\documentclass[twocolumn,tighten,times]{aastex63}

\usepackage{graphics,epsf}
\usepackage{amsmath}                % American Mathematical Society package
\usepackage{amsfonts}               % American Mathematical Society fonts
\usepackage{amssymb}                % American Mathematical Society symbol
\usepackage{epsfig}
\usepackage{epstopdf}

\usepackage{xcolor}
\definecolor{redak}{rgb}{1.0,0.0,0.0}
\def \msyr{~\rm{M_{\odot}}~\rm{yr^{-1}}}
\def \cm{~\rm{cm}}
\def \s{~\rm{s}}
\def \km{~\rm{km}}

\def \K{~\rm{K}}
\def \g{~\rm{g}}
\def \G{~\rm{G}}
\def \AU{~\rm{AU}}

\def \yr{~\rm{yr}}

\def \kpc{~\rm{kpc}}
\def \etc{$\eta$~Car}
\def \days{~\rm{days}}

\def \keV{~\rm{keV}}

\def \rmModot{~\rm{M_{\sun}}}

\begin{document}

\title{\large{The X-ray Properties of Eta Carinae During its 2020 X-ray Minimum}}

\author[0000-0002-7840-0181]{Amit Kashi}
\affil{Department of Physics, Ariel University, Ariel, POB 3, 4070000, Israel; \url{kashi@ariel.ac.il}}
%\email{kashi@ariel.ac.il}
\author[0000-0002-7939-377X]{David A. Principe}
\affil{Massachusetts Institute of Technology, Kavli Institute for Astrophysics and Space Research, Cambridge, MA, USA}
%\email{principe@space.mit.edu}
\author[0000-0003-0375-8987]{Noam Soker}
\affil{Department of Physics, Technion -- Israel Institute of Technology, Haifa 3200003, Israel}
\affil{Guangdong Technion Israel Institute of Technology, Guangdong Province, Shantou 515069, China}
%\email{soker@physics.ac.il}
\author[0000-0002-3138-8250]{Joel H. Kastner}
\affil{Center for Imaging Science, School of Physics \& Astronomy, and Laboratory for Multiwavelength Astrophysics, Rochester Institute of Technology, Rochester, NY, USA}
%\email{jhkpci@cis.rit.edu}

\shorttitle{Eta Car 2020 X-ray minimum} 
\shortauthors{A. Kashi et al.}

\begin{abstract}
The massive binary system Eta Carinae is characterized by intense colliding winds that form shocks and emit X-rays. The system is highly eccentric ($e\simeq0.9$), resulting in modulated X-ray emission during its 5.54 year orbit. The X-ray flux increases in the months prior to periastron passage, exhibiting strong flares, then rapidly declines to a flat minimum lasting a few weeks, followed by a gradual recovery. We present Neutron Star Interior Composition Explorer (\textit{NICER}) telescope spectra obtained before, during, and after the 2020 X-ray minimum, and perform spectral analysis to establish the temporal behavior of X-ray flux and X-ray-absorbing column density ($N_{\rm H}(t)$) for the 2--10 keV and 5--10 keV energy ranges. The latter range is dominated by the stellar wind collision region and, therefore, these spectral parameters --- in particular, $N_{\rm H}(t)$ --- serves as a potentially stringent constraint on the binary orientation. We compare the observed $N_{\rm H}(t)$ results to the behavior predicted by a simple geometrical model in an attempt to ascertain which star is closer to us at periastron: the more massive primary ($\omega \simeq 240$--$270^\circ$), or the secondary ($\omega \simeq 90^\circ$). We find that the variations in column density, both far from periastron and around periastron passage, support the latter configuration ($\omega \simeq 90^\circ$). The 2020 X-ray minimum showed the fastest recovery among the last five minima, providing additional evidence for a recent weakening of the primary star's wind.
\end{abstract}


% ==========================================================
\section{INTRODUCTION}
\label{sec:intro}
% ==========================================================

In early 2020 the binary system Eta Carinae (\etc) underwent another periastron passage in its ${P = 2023}$ day orbit.
The system includes two very massive stars, the primary  and the hotter secondary. 
The primary star is a very massive star with mass $M_1=120$--$170 \rmModot$ \citep{Hillieretal2001, DavidsonHumphreys2012, KashiSoker2010, KashiSoker2016}, that may be in its Luminous Blue Variable (LBV) stage (though it differs from classical LBVs in a few aspects, see \citealt{HillierAllen1992}). The secondary star is a hot and evolved star \citep{Verneretal2005, Mehneretal2010} with a mass of $M_2=30$--$80 \rmModot$ \citep{KashiSoker2010, KashiSoker2016}, that may be in its Wolf-Rayet stage \citep[e.g.,][]{Hiraietal2021}.
The system is a colliding wind binary \citep{Damineli1996, Pittardetal1998} with large eccentricity $e \simeq 0.9$ \citep{Davidsonetal2017}, resulting in large differences in the wind collision interface between periastron and apastron.
The distance to the system has been determined to be in the range $2.3 \kpc$ \citep{Smith2006}
to $2.6 \kpc$ \citep{Davidsonetal2018a}.
The system is famous for its energetic eruptions in the nineteenth century \citep{DavidsonHumphreys2012}, the great eruption (GE; 1837.9--$\sim$1858) and the lesser eruption (LE; 1887.3--1895.3). These eruptions overpowered the Eddington limit and ejected a significant portion of the primary's stellar atmosphere, forming what we know today as the Homunculus nebula \citep{DavidsonHumphreys2012}.
 
The primary wind has a larger momentum than that of the secondary, with mass loss rate $\dot M_1 \simeq 3$--$10 \times 10^{-4} \msyr$ \citep{DavidsonHumphreys2012,Clementeletal2014,Kashi2017,Kashi2019}.
The weaker secondary wind carves a conical cavity in the dense primary wind with direction and shape that changes with the orbit. This cavity wraps around the primary close to periastron passage, when the orbital velocity increases and becomes comparable to the primary wind velocity \citep{Hamaguchietal2007,Parkinetal2011,KashiSoker2009a,Maduraetal2012}.

Being such a unique system at a relatively close distance, each $\eta$~Car periastron passage is an event of great interest and is monitored from both ground \citep{DuncanWhite2003,vanGenderenSterken2004,Whitelocketal2004,Abrahametal2005b,Daminelietal2008a,Daminelietal2008b,FernandezLajusetal2010,Teodoroetal2012,Teodoroetal2016} and space \citep{PittardCorcoran2002, Martinetal2006, Corcoran2005, Corcoranetal2010, Corcoranetal2017, Hamaguchietal2007, Hamaguchietal2014a, Hamaguchietal2014b, Henleyetal2008, Davidsonetal2015, Mehneretal2015}.
Prior to periastron passage, $\eta$ Car's spectral lines, which are key probes of the dynamics of the two stars and their winds, change their profiles, some dramatically. This behaviour of the lines as well as rapid variations in various bands, from IR to X-ray, around periastron passages has earned the name `the spectroscopic event' \citep{DavidsonHumphreys2012}.

The spectroscopic events observed over multiple periastrons have varied considerably, with a trend of becoming shorter and less intense. This may be due either to a change of state in the primary wind \citep{Mehneretal2010,Davidsonetal2018b,Kashietal2016}, or related to the dissipation of the surrounding Homunculus Nebula, at least along our line of sight \citep{Daminelietal2019,Mehneretal2019}.

The spectroscopic events of $\eta$~Car should yield insight into the binary's geometry and, more generally, the physics of stellar wind interactions.
However, there is still disagreement on the orientation.
The inclination of the binary was initially assumed to be the same as the orientation of the Homunculus nebula, $i \simeq 41^\circ$ 
\citep{Davidsonetal2001}, but the direction of motion was ambiguous.
\cite{Maduraetal2012} and \cite{Teodoroetal2016} deduced the inclination to be $i=130^\circ$--$145^\circ$ and $i=135^\circ$--$153^\circ$, respectively, suggesting a direction of orbital motion opposite to that adopted by \citet{Davidsonetal2001}. The foregoing range in inferred binary inclination is narrow enough not to pose any difficulties for purposes of interpreting observational data; nor is the direction of binary motion important, for such purposes.

Unlike the inclination, the argument of periapsis $\omega$, for which different values have been obtained, has significant implications for understanding the behavior of $\eta$~Car near periastron passages.
A value of $\omega=90^\circ$ implies that the secondary star (which launches the fast wind) is closest to us at periastron and furthest at apastron, while a value of
$\omega=270^\circ$ implies that the primary star (which launches the slow and dense wind) is closest to us at periastron and furthest at apastron.
A number of studies use fitting of spectral line profiles to claim the orientation is $\omega\approx270^\circ$
\citep[e.g.][]{Nielsenetal2007, Henleyetal2008, Daminelietal2008b, Richardsonetal2015}.
\cite{Maduraetal2012} ran Smooth Particle Hydrodynamics (SPH) simulations of the colliding winds and, by matching the simulation results to line spectro-imaging data, determined an orientation within the range $\omega=240^\circ$--$285^\circ$.
\cite{Weigeltetal2016} claimed to have identified a fan-shaped structure in the wind. They combined their claim with the simulations of \cite{Maduraetal2012} to support their preferred orientation of $\omega\approx240$--$270^\circ$. However, in \cite{KashiSoker2018} we showed that the fan probably does not exist, and even if so it does not support the orientation that \cite{Weigeltetal2016} argued for.
Recently, \cite{Grantetal2020} found a slightly higher eccentricity $e=0.91$, and determined $\omega=241^\circ$ by fitting Balmer lines with Gaussian components and then fitting a Keplerian model.

Fitting an orbital orientation from spectral lines depends on assumptions as to where these lines are emitted or where they are absorbed. Attributing the lines to different locations can result in the opposite solution for $\omega$, i.e., $\omega\approx90^\circ$
\citep{KashiSoker2007,KashiSoker2008a,KashiSoker2016, Kashietal2011}.
\cite{KashiSoker2018} showed that the orientation with $\omega=90^\circ$ can explain the absence of a mass segment in the torus that was ejected during the GE, as described by \cite{Smithetal2018b}.
\cite{Abrahametal2005a}, \cite{Falcetaetal2005} and \citep{AbrahamFalceta2007} also proposed an orientation with $\omega\approx60$--$90^\circ$, but their model invokes a shell ejection event to explain the X-ray behavior of $\eta$~Car during the spectroscopic event and is hence quite different from the model of \cite{KashiSoker2008a} (discussed below).

X-ray observations of $\eta$~Car have also been used to derive the orientation. \cite{Okazakietal2008} ran SPH simulations of the colliding winds and, based on fits to the X-ray luminosity, claim $\omega=243^\circ$.
\cite{Parkinetal2009} built a model to fit the X-ray light curve and derived $\omega=270^\circ$--$300^\circ$. However,
\cite{KashiSoker2009a} showed that the X-ray light curve is not a strong indicator of the binary orientation.
They suggested that the hydrogen column density to the hot X-ray emitting gas is a better observable for purposes of differentiating between the different proposed orientations, because the hot X-ray emitting gas arises from within the post-shock secondary wind located between the two stars.

\cite{KashiSoker2009a} analyzed the expected X-ray flux from $\eta$~Car considering a specific accretion model.
Their model includes the slow-dense primary and fast secondary winds, an approximation for the colliding wind interface, and analytic expressions for the thickness of the regions that include the post-shock primary and secondary winds. 
They also included in their model the rotation angle that the conical shell forms with the line connecting the two stars as the secondary approaches periastron passage and the orbital velocity becomes significant.
They calculated the expected time dependence of the hydrogen column density $N_{\rm{H}}(t)$ from different directions, demonstrating that it can serve as an indicator of the binary orientation.
They found that for orientations in which the primary is closer to the observer at periastron passage ($\omega \simeq 270^\circ$), the primary wind absorbs so much of the colliding wind X-ray flux that no X-rays should be detected.
They could also explain the observed X-ray emission measure with their preferred orientation of $\omega=90^\circ$.

Close to periastron passage the X-ray flux decreases, which makes observations more challenging. The central binary (the shocked secondary wind close to the apex, located between the two stars) that is responsible for the  hard X-ray emission dims.
This dimming can be explained as due either to lower emission from the wind-collision region and/or a very large increase in $N_{\rm H}$ toward the wind-collision region.
\cite{Hamaguchietal2007} demonstrate the existence of an emission component referred to as a 'central constant emission component' (CCE) which can significantly contribute to the X-ray spectrum during periastron. This component lies within $\sim$1 arcsecond of the central binary and arises from a larger (but unresolved) region surrounding the binary system and, due to its larger plasma volume, does not significantly vary in time. \cite{Hamaguchietal2007} find that this component can be fit with an $\sim$ 1.1 keV plasma and an $N_{\rm H} \sim 5 \times 10^{22} \cm^{-2}$. This component is further discussed in \cite{Hamaguchietal2014a,Hamaguchietal2014b,Hamaguchietal2016}.
During periastron, it is unclear to what extent this CCE component dominates the entire X-ray spectrum.
\cite{Hamaguchietal2014a,Hamaguchietal2014b,Hamaguchietal2016} analyzed X-ray observations taken by \textit{Chandra}, \textit{XMM-Newton} and \textit{NuSTAR} telescopes, and found that the hydrogen column density to the central binary near periastron may be as large as $N_{\rm H} \approx 10^{24} \cm^{-2}$.
The smooth variations of $N_{\rm H}$ as a function of time before, during and after periastron could suggest that the central binary is not completely obscured and as a result, $N_{\rm H}$ measurements could at least partially probe the central binary system. Such smooth variations were demonstrated in the 2--10 keV range with the Rossi X-ray Timing Explorer (\textit{RXTE}) in \cite{Ishibashietal1999}.

In early 2020, the $\eta$~Car system underwent another periastron passage.
The X-ray light curve was observed by the Neutron Star Interior Composition Explorer (\textit{NICER}) observatory \citep{Corcoranetal2020,Espinoza-Galeasetal2020}.
The X-ray light curve reached a minimum on Feb 14, 2020 \citep{Corcoranetal2020,Espinoza-Galeasetal2020}.
The exact date of periastron is uncertain and may begin about a week or two around that date; we will hence refer to it as the 2020.1 periastron passage.
The light curve shows similar properties to that of previous X-ray minima \citep{Corcoranetal2017},
but with a somewhat earlier exit from the minimum. 
The X-ray flux increases in the months preceding the periastron passage, with strong flares superimposed on top of the smooth $\sim 1/r$ increase. These flares are most likely associated with clumps in the wind \citep{MoffatCorcoran2009}, and were also observed prior to the 2020 minimum \citep{Corcoranetal2019}.
After showing a few strong X-ray flares the light curve sharply declines into a non-zero minimum which lasts for several weeks, then slowly recovers to a quiescence value.
\cite{Hamaguchietal2020} also report bright non-thermal X-ray emission in \textit{NuSTAR} observations taken during the recovery from the 2020 X-ray minimum.

In this paper, we model X-ray data for the 2020.1 periastron passage, obtained with the \textit{NICER} X-ray telescope, with the goal of explaining the variation of the X-ray luminosity $L_{\rm X}$ and of the column density $N_{\rm H}$ in the frame of the accretion model for the spectroscopic event. 
According to the accretion model the X-ray minimum is not caused by an absorption. Rather, the secondary star accretes mass from the primary stellar wind for several weeks near periastron passages \citep{Soker2005, Soker2007, SokerBehar2006, Akashietal2006, KashiSoker2007, KashiSoker2008b, KashiSoker2009c}. This accretion suppresses the secondary star's wind. 
Since the source of the hard X-ray emission is the post-shock fast, $v_2 \simeq 3000 \km \s^{-1}$, secondary wind, the X-ray luminosity displays a minimum with a duration of several weeks.
Hydrodynamic simulations show that indeed the secondary star accretes mass near periastron, in a process that suffers instabilities \citep{Akashietal2013, Kashi2017, Kashi2019}.
The accretion model can account for several other observed properties, such as the orbital variations of  numerous lines \citep{KashiSoker2007,KashiSoker2008a, KashiSoker2009a,KashiSoker2009b,KashiSoker2016,KashiSoker2018}, the infrared light curve \citep{KashiSoker2008b}, the X-ray light curve and emission measure \citep{SokerBehar2006, Akashietal2006, KashiSoker2009a}, the timing of the peaks in the light curve of the GE and LE \citep{KashiSoker2010}, the very fast velocities of gas ejected during the GE \citep{AkashiKashi2020}, and more.

\cite{Navareteetal2020} observed optical lines during the 2020.1 spectroscopic event and concluded that the circumstellar ejecta is dissipating, but the primary does not change significantly. Similar conclusions were reported by \cite{Mehneretal2019} and \cite{Daminelietal2019}.
On the other hand, long term observations of optical lines $\eta$~Car lead \cite{Davidsonetal2005} to suggest there is a decrease in the mass-loss rate from the primary star, referred to as a `change of state'.
This effect was obtained in numerical simulations of the recovery of $\eta$~Car from the great eruption \citep{Kashietal2016}.
Further indication for the change came from comparison of UV lines emission at similar orbital phases separated by two orbital revolutions, at positions far from periastron passage \citep{Davidsonetal2018b}.
Here we show that the 2020 X-ray minimum provides further evidence for a recent decrease in the primary's mass loss rate.

In section \ref{sec:theory_NH} we describe the expected general behavior of $N_{\rm H}$ around periastron passage. We then describe the observations (section \ref{sec:obs}) and the X-ray light curve (section \ref{sec:X-Ray-LC}). 
We return to the accretion model and instabilities in our analysis of the new observations in section \ref{sec:NH}. We summarize our results in section \ref{sec:summary}.

% ==========================================================
\section{Theoretical considerations for the X-ray column density}
\label{sec:theory_NH}
% ========================================================

\begin{figure} % [ht]
\centering
\includegraphics[trim=0.8cm 14cm 4.0cm 2.8cm ,clip, scale=0.53]{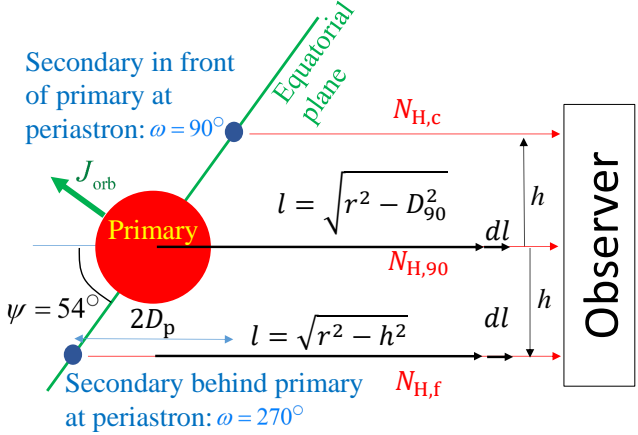}  %l b r t
% \vspace*{-2.5cm}
%\hskip -0.5 cm
\begin{singlespace}
\caption{Schematic drawing of the geometry of the binary system in the plane that contains the center of the primary star, the secondary at periastron, and the observer. We examine two opposite orientations where during periastron the secondary star is either at the closest  ($\omega \simeq 90^\circ$) or furthest point from us  ($\omega \simeq 270^\circ$).  We draw both alternative orientations on the figure by indicating the two alternative periastron locations of the secondary star. The green arrow indicates the orbital angular momentum axis of the binary system. The dashed horizontal line indicates that the integration is not in the plane of the figure, but at a distance of $D_{90}$ from the plane (see text). 
}
\end{singlespace}
\label{fig:Schematic}
\end{figure}

We derive the column density for three cases according to the position of the secondary star with respect to the primary star and the observer. These cases are indicated as $N_{\rm H,90}$, $N_{\rm H,c}$, and $N_{\rm H,f}$ in Figure \ref{fig:Schematic}. One case ($N_{\rm H,90}$) is relevant to the two opposite orientations, while each of the two other cases is relevant to a different, specific orientation.
In one orientation ($N_{\rm H,c}$) the secondary is closest to us at periastron ($\omega \simeq 90^\circ$), while in the other ($N_{\rm H,f}$) the secondary star is furthest from us at periastron ($\omega \simeq 270^\circ$). For the inclination of the orbit we adopt $i=144^\circ$, based on results in \cite{Maduraetal2012} and
\cite{Teodoroetal2016}; the adopted value represents the middle of the range determined by \cite{Teodoroetal2016}.
The inclination is defined as the angle from the line of sight to the angular momentum of the binary system (along a line perpendicular to the orbital plane). In Figure \ref{fig:Schematic} we indicate the angle between the line of sight and the orbital plane. According to \cite{Teodoroetal2016}, in the `secondary furthest orientation’ at periastron the secondary is not precisely behind the primary, but there are other uncertainties that are larger. One such uncertainty is the X-ray emitting zone (see below). Another uncertainty is the exact density structure of the primary stellar wind, i.e., its exact mass loss rate, how clumpy it is, and how the interaction zone of the two winds behave (see \citealt{KashiSoker2008a}).  In Figure \ref{fig:Schematic} we indicate with a solid black arrow the integration line for the column density in the `secondary furthest orientation', $N_{\rm H,f}$. This line is in the plane of the diagram. For the `secondary closest orientation' column density $N_{\rm H,c}$, we will obtain a similar expression for the line of integration (solid red arrow at top of diagram), but with different integration boundaries. 
% FFFFFFFFFFFFFFFFFFFFFFFFFFFFFFFFFFFFFFFFFFFFFFFFFF
% FFFFFFFFFFFFFFFFFFFFFFFFFFFFFFFFFFFFFFFFFFFFFFFFFF 
    
We define the angle $\psi=i-\pi/2 = 54^\circ$ as indicated in Figure \ref{fig:Schematic}.
These integration lines are at a distance of $h=a(1-e)\sin \psi = 1.35 \AU$ from the line of sight to the center of the primary star,   % 1.3462
where we take $a=16.64 \AU$ and $e=0.9$ for the semi-major axis and eccentricity, respectively. For a period of 2023 days, this value of the semi-major axis implies a total binary mass of $M_1+M_2= 150 M_\odot$. 
Note that we took here what is usually referred to as the conventional model for the masses of $\eta$ Car, but as will be seen from the following equations (that are calibrated for the conventional model) the values of the column density only slightly change for the high-mass model, for which $a=19.73 \AU$ and $M_1+M_2= 250 M_\odot$ (see table 1 in \citealt{Kashi2017}).
In addition to $N_{\rm H,c}$ and $N_{\rm H,f}$, we also calculate the column density from the secondary star to the observer when the primary-secondary line is perpendicular to the line of sight, $N_{\rm H,90}$. This line of integration lies at a distance of $D_{90}=a(1-e^2) = 3.16 \AU$ above or below the plane of Figure \ref{fig:Schematic}. %3.1616
   
Expressions for the foregoing column densities are obtained as follows. The proton number density of the primary wind is 
\begin{equation}
n_{\rm H} = \frac {X \dot M_1}{4 \pi r^2 m_{\rm H} v_1} \equiv \Lambda r^{-2}, 
\label{eq:nHwind}
\end{equation}
where $X$ is the hydrogen mass fraction, $v_1$ is the velocity of the primary wind, and $\dot M_1$ is the mass loss rate into the primary wind, and the second equality defines  the parameter
\begin{equation}
\begin{split}
\Lambda &= 6.01 \times10^{23} \left( \frac {X}{0.5} \right)
\left(\frac{\dot M_1}{3 \times 10^{-4} \rmModot \yr^{-1}} \right) \\
& \times \left( \frac {v_1}{500 \km \s^{-1}} \right)^{-1}  \cm^{-2} \AU,  %6.013427342 +23
\end{split}
\label{eq:Lambda}
\end{equation}
where the unusual choice of units facilitates later scaling.

In this calculation we assume that the bulk of the observed 2--10 keV X-ray emission from $\eta$~Car arises very close to the secondary star. The emission should in fact be generated at some distance between the stars, which will reduce the difference in distance between the two orientation cases at periastron, but on the other hand will bring the X-ray emitting gas deeper into the dense primary wind. The latter effect is larger, and as a consequence this model may underestimate the difference in column densities between the two orientations. 
Given these uncertainties, the calculation we present here is an approximation. Nonetheless, we think is adequate for our purposes, i.e., interpreting the observed X-ray flux and absorption in terms of the binary orbital parameters.

We can now calculate the column density for the three cases illustrated in Figure \ref{fig:Schematic}. As  noted,
the first case, column density $N_{\rm H,c}$, is for the case where the secondary closer to the observer at periastron, while the second, $N_{\rm H,f}$, is for the secondary behind the primary wind at periastron.
The third value of the column density, $N_{\rm H,90}$, is the same for both orbit orientations in the simple form presented in Figure \ref{fig:Schematic}, and this third case applies when the secondary-primary line is perpendicular to the line of sight. At that time the distance between the secondary and primary is $D_{90} = a (1-e^2) = 3.16 \AU $
%3.1616
for the parameters we adopt here.
We can obtain these column densities as follows. For the perpendicular phase we have
\begin{equation}
\begin{split}
N_{\rm H,90} &= \int^{\infty}_0 n_{\rm H} dl = \int^{\infty}_0 \Lambda \frac{dl}{l^2 + D^2_{90}}  = \frac{\Lambda}{D_{90}} \frac {\pi}{2} \\
&= 2.99 \times 10^{23} \left( \frac{\Lambda}{6.01 \times10^{23}  \cm^{-2} \AU} \right) \\
&\times \left( \frac{D_{90}}{3.16 \AU} \right) ^{-1} \cm^{-2}.  %2.9877
\end{split}
\label{eq:NH90}
\end{equation}
Using the geometry as presented in Figure \ref{fig:Schematic} we find that $D_{\rm p}=a(1-e) \cos \psi$, and so $D_{\rm p}/h=\cot \psi=0.73$. %0.72654
We then  obtain
\begin{equation}
\begin{split}
N_{\rm H,c} &= \int^{\infty}_{D_{\rm p}} n_{\rm H} dl = \int^{\infty}_{D_{\rm p}} \Lambda \frac{dl}{l^2 + h^2}  \\
& = \frac{\Lambda}{h} \left( \frac {\pi}{2} - \tan^{-1} \frac{D_{\rm p}}{h} \right) = \frac{\Lambda}{h} \psi  = \frac{\Lambda}{r_{\rm{p}}} \frac{\psi}{\sin \psi} \\
&= 4.21 \times 10^{23} \left( \frac{\Lambda}{6.01 \times10^{23}  \cm^{-2} \AU} \right) \\
&\times \left( \frac{r_{\rm{p}}}{1.66 \AU} \right) ^{-1}  \cm^{-2} ,   %4.2099e23
\end{split}
\label{eq:NHc}
\end{equation}
where $r_{\rm{p}}=1(1-e) =1.66 \AU$ is the periastron distance.
Similarly, 
\begin{equation}
\begin{split}
N_{\rm H,f} &= \int^{\infty}_{-D_{\rm p}} n_{\rm H} dl = \int^{\infty}_{-D_{\rm p}} \Lambda \frac{dl}{l^2 + h^2}  
\\& = \frac{\Lambda}{h} \left( \frac {\pi}{2} + \tan^{-1} \frac{D_{\rm p}}{h} \right) = \frac{\Lambda}{h} (\pi+ \psi)  = \frac{\Lambda}{r_{\rm{p}}} \frac{\pi+\psi}{\sin \psi} \\
& = 9.82 \times 10^{23} \left( \frac{\Lambda}{6.01 \times10^{23}  \cm^{-2} \AU} \right) \\
&\times \left( \frac{r_{\rm{p}}}{1.66 \AU} \right) ^{-1}  \cm^{-2} ,   %9.82329
\end{split}
\label{eq:NHf}
\end{equation} 
These expressions show that if the distance from primary to X-ray emitting region is smaller than the distance from primary to secondary
(i.e., X-rays are generated in between the two stars),
then the column density is larger than the calibrated values in equations (\ref{eq:NHc}) and (\ref{eq:NHf}).

For later analysis the predicted ratios of these column densities are
\begin{equation}
\frac{N_{\rm H,c}}{N_{\rm H,90}} = 1.41 \left(\frac{1+e}{1.9}\right)\quad {\rm and}
\quad 
\frac{N_{\rm H,f}}{N_{\rm H,90}} = 3.29 \left(\frac{1+e}{1.9}\right),
\label{eq:ratios}
\end{equation}  % 1.409  3.2879 
for the closest-orientation and furthest-orientation, respectively. 
Note that the above ratios do not depend on $\Lambda$, and even do not depend on $a$.
They only depend on $(1+e)$, and as the eccentricity is a well constraint variable, if we take $e\simeq0.88$--$0.92$ these ratios vary only by 1\%. We note again that this calculation assumes a smooth primary wind that fills the entire volume.

% ==========================================================
\section{Observations}
\label{sec:obs}
% ==========================================================

\begin{figure*}
\includegraphics[trim= 0.0cm 0.0cm 0.0cm 1.5cm,clip=true,width=0.99\textwidth]{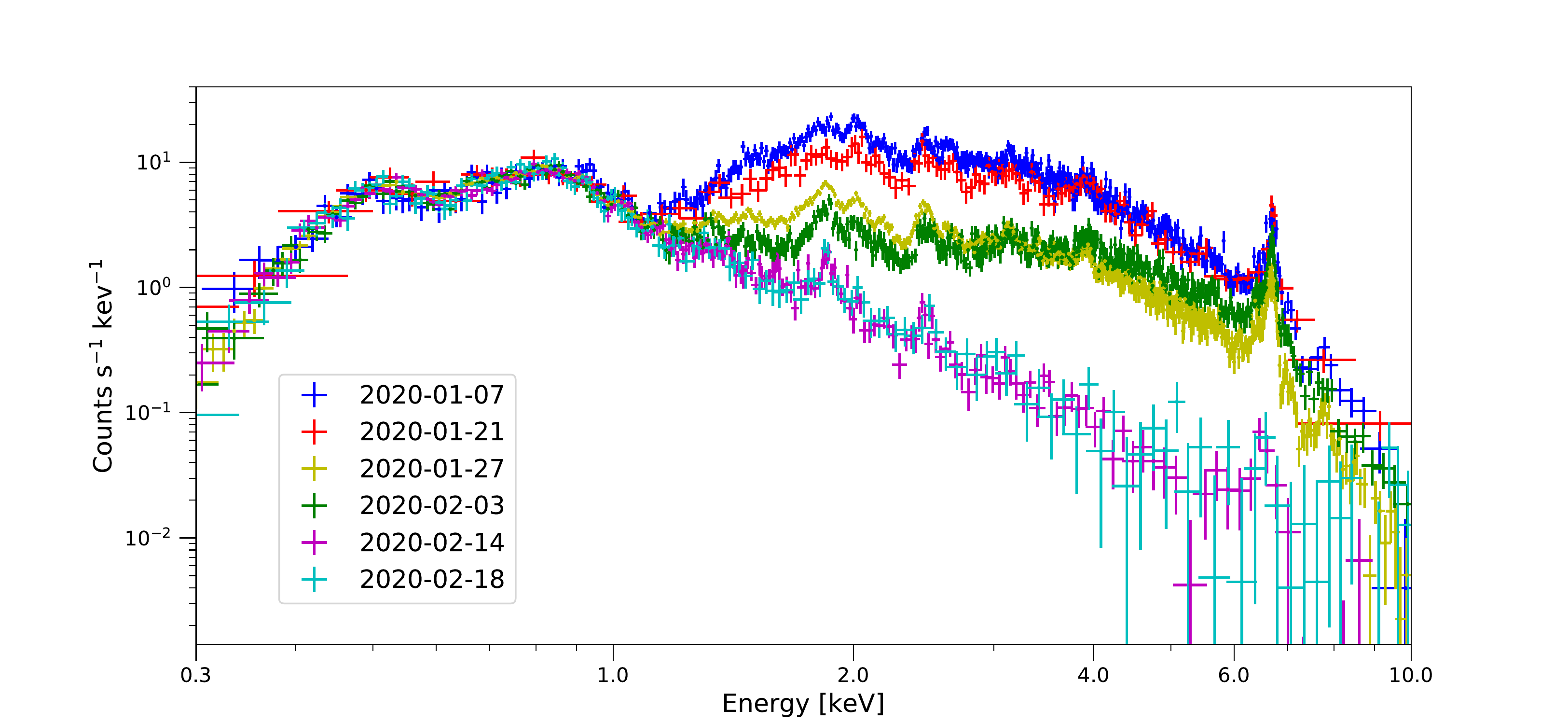}
\begin{singlespace}
\caption{
\textit{NICER} X-ray spectra of $\eta$ Car across the 2020 periastron passage grouped to require a minimum SNR$=$6. Observations of the flare are displayed on January 7, 21, 27, and February 3. \textit{NICER} spectra during the X-ray minimum are displayed for February 14 and 18. The extended X-ray emitting regions surrounding $\eta$~Car contribute primarily at energies $<$ approximately 2 keV where little variability is observed. The 2--10 keV emission is associated with previously shocked gas farther from the central stars. Emission in the 5--10 keV range comes from the apex of the colliding winds located between the two stars but closer to the secondary.}
\end{singlespace}
\label{fig:xrayspectra}
\end{figure*}

Observations of $\eta$ Car were obtained with the Neutron Star Interior Composition Explorer (\textit{NICER}), under \textit{NICER} Guest Observer programs 1110 (PI: K. Gendreau), 2612 (PI: M. Corcoran), and 3651 (PI: D. Espinoza-Galeas). \textit{NICER} is an X-ray telescope attached to the International Space Station. \textit{NICER}'s large effective area, broad band pass, moderate spectral resolution and 30 arcmin$^{2}$ field of view provide the capability of collecting spatially unresolved observations of X-ray emitting regions surrounding $\eta$~Car. The Point Spread Function (PSF) is approximately the entire field of view ($R\simeq3.1$ arcmin).
Archival observations of $\eta$~Car by \textit{NICER} were obtained for observing dates between 2017-07-20 and 2020-07-24. Archival data were reduced using the HEASoft package (v 6.27.2) available from the High Energy Astrophysics Archive Research Center (HEASARC). The \textit{NICER} Heasoft tool \texttt{nicerl2}\footnote{\url{https://heasarc.gsfc.nasa.gov/lheasoft/ftools/headas/nicerl2.html}}
was used to create cleaned level 2 event files with up-to-date gain calibrations (\textit{NICER} CALDB version 20200722). Source and background spectral files for each observation were extracted using the \textit{NICER} background estimator tool \texttt{nibackgen3c50}\footnote{\url{https://heasarc.gsfc.nasa.gov/docs/nicer/tools/nicer_bkg_est_tools.html}} (v5). Due to its flux variability, $\eta$~Car occasionally reaches the detection threshold of \textit{NICER} in a given observation (Figure \ref{fig:xrayspectra}). To prevent adding unnecessary background counts to our source spectra, we chose to remove portions of an observation if the good time intervals indicated an exposure of less than 100 seconds. These very short exposures likely include very few source counts but can have high background levels. A maximum good time interval (GTI) time of 3000 seconds and a maximum net high background rate value of 0.05 counts s$^{-1}$ were chosen as input parameters when generating background spectra for the occasionally faint source $\eta$~Car. Of the $\approx 290$ \textit{NICER} observations with non-zero exposure times in our date range, $\approx 50$ observations did not meet the above criteria for background selection were discarded.

The \textit{NICER} instrument is unable to produce spatially resolved spectra for the $\eta$~Car region; all X-ray emission in the 30 arcmin$^{2}$ field of view (FOV) surrounding $\eta$~Car is included in these spectra. Thus, as discussed in \cite{Hamaguchietal2007} for the case of \textit{XMM-Newton} observations of $\eta$~Car, several different X-ray emission components associated with $\eta$~Car can be observed in the \textit{NICER} spectra. In particular, the outer ejecta regions and the Homunculus Nebula both contribute significantly. Since we cannot remove these contributions to the \textit{NICER} spectra of $\eta$~Car, we instead focus on relative changes in the X-ray spectrum due to the variability of the $\eta$~Car central binary as the system first approached and then recovered from periastron. We assume that on timescales of days to months, no variability is induced by these external X-ray components in the 2--10~keV range where we perform our spectral fits.

We checked the field of view (FOV) of \textit{NICER} in order to exclude potential contamination from other X-ray sources.
For that we used the catalogue of X-rays sources from the \textit{CHANDRA} Carina Complex Project \citep{Broosetal2011} to identify other sources that may contribute to the X-ray flux.
The FOV includes $\simeq400$ additional sources, the vast majority of which are young stellar sources pre-main sequence stars that have median X-ray energies $<2$~keV in the 0.5--8~keV range. Individually their fluxes are negligible compared to the extremely bright $\eta$~Car in the 2--10~keV and 5--10~keV ranges.
The most notable X-ray point source that falls within the field of view of the \textit{NICER} $\eta$~Car observations is HDE~303308 \citep[a spectroscopic binary O4.5V star;][]{Sotaetal2014}.
The combined 2--8~keV flux of all these $\sim 400$ field X-ray sources is $\simeq1.5\times 10^{-12}~\rm{erg~s^{-1}~cm^{-2}}$ which is lower than all observations which we include in our analysis of $N_{\rm H}$ presented here.
Moreover, these sources are distributed across the \textit{NICER} field of view and hence their X-ray detection rates are suppressed relative to the on-axis $\eta$~Car source
Only one observation during the X-ray minimum (on JD$=$2458895.7) has a 2--10~keV flux smaller that $3\times 10^{-12}~\rm{erg~s^{-1}~cm^{-2}}$, and hence might include a contribution from the field X-ray sources.

Fits to the resulting \textit{NICER} spectra were obtained via the Sherpa\footnote{\url{https://cxc.cfa.harvard.edu/sherpa/}} X-ray spectral fitting package (v 4.12).
We constructed a python script to automate the fitting process for the 242 \textit{NICER} observations.
Each spectrum was background subtracted and grouped to require a signal to noise ratio (SNR) of at least 3 per bin.
Fits were performed for each spectrum between two energy ranges, 2--10 keV and 5--10 keV (see section \ref{sec:NH}), using the Nelder-Mead Simplex optimization method with the $\chi^{2}$ Gehrels statistic.
Our tests showed that grouping the spectra by counts per bin (e.g.,
grouping by 5 or 15 counts per bin) rather than by signal-to-noise ratio
did not consistently improve the fits.

Following a similar procedure to that discussed in \cite{Hamaguchietal2007}, we fit each spectrum with an absorbed \texttt{{{VAPEC}}} thermal equilibrium model \citep[using \texttt{wabs} for the absoption component of the model;][]{MorrisonMcCammon1983} and two Gaussians at the positions of the Fe K$\alpha$ and K$\beta$ lines (6.4 keV and 7.1 keV, respectively). An additional Gaussian component was used in the 2--10 keV fits to model the S {\sc xv} line at $\sim 2.5 \keV$; this component is associated with the Homoculus Nebula \citep{Hamaguchietal2007}. The Fe abundance remained free between 0.0 and 1.0 times solar in the fits, while the other metals were frozen to their solar values \citep{Hillieretal2001}. The ratio of the Fe K$\alpha$ flux to Fe K$\beta$ was set to 11.3\% \citep{Hamaguchietal2007, Yamaguchietal2014}.

Spectral fitting results, including the observed X-ray fluxes calculated for both the 2--10 keV and 5--10 keV fits and $\chi^2_{\rm red}$ values, are listed in 
Tables 1 and 2.
%Tables \ref{table:210} and \ref{table:510}.
All fit parameters are reported with their 90\% confidence intervals. The fluxes reported for the 2--10~keV fits do not include the S {\sc xv} Gaussian component.
Emission measure (EM) is reported from the fit normalization parameter assuming a distance of $2.6 \kpc$. Large uncertainties are present in the 5--10 keV best-fit parameters as expected for a source with low flux in this energy range. For both the 2--10 keV and 5--10 keV fits, we do not report values or errors where the best-fit model did not converge on a particular parameter value (i.e., a maxima or minima boundary was hit).
\renewcommand{\arraystretch}{0.98}
\begin{table*}[]
\begin{center}
%\centering
\begin{singlespace}
\caption{Best-fit model parameters to the 2--10 keV \textit{NICER} X-ray spectra of Eta Car between July 2017 and July 2020 and their 90\% confidence values.}
\end{singlespace}
\begin{tabular}{c c c c c c c c c}
\hline\hline
Date & Obs.\,ID. & $N_\mathrm{H}$ & $kT$  & Fe & log(EM) & Flux $\times10^{-11}$ & $\chi^2_{\rm red}$\\
$[$UT$]$ &  & $[10^{22}$ cm$^{-2}]$ & $[$keV$]$  & $[$Z$_{\odot}]$ & $[$cm$^{-3}]$ & $[$erg s$^{-1}$ cm$^{-2}]$ & \\
\hline
2017-07-20T02:59:20 & 1110010101 & 2.7$^{+0.9}_{-0.9}$ & 3.6$^{+1.6}_{-0.9}$ & 0.67$^{+0.37}_{-0.33}$ & 57.70$^{+0.12}_{-0.13}$ & 4.2$^{+1.0}_{-1.0}$ &0.67\\
2017-07-21T00:36:20 & 1110010102 & 3.0$^{+0.6}_{-0.7}$ & 3.2$^{+0.9}_{-0.5}$ & 0.76$^{+0.28}_{-0.25}$ & 57.79$^{+0.08}_{-0.10}$ & 4.5$^{+1.4}_{-1.1}$ &0.54\\
2017-07-22T04:23:20 & 1110010103 & 3.3$^{+0.6}_{-0.7}$ & 2.9$^{+0.6}_{-0.4}$ & 0.60$^{+0.28}_{-0.25}$ & 57.84$^{+0.09}_{-0.09}$ & 4.2$^{+0.6}_{-0.6}$ &0.61\\
2017-07-24T10:25:00 & 1110010105 & 2.8$^{+1.6}_{-1.3}$ & 3.4$^{+2.6}_{-1.2}$ & 0.79$^{+0.66}_{-0.50}$ & 57.72$^{+0.22}_{-0.18}$ & 4.0$^{+3.5}_{-2.8}$ &0.62\\
2017-10-05T02:11:20 & 1110010106 & 3.6$^{+0.8}_{-0.7}$ & 2.5$^{+0.6}_{-0.5}$ & 0.54$^{+0.35}_{-0.29}$ & 57.92$^{+0.12}_{-0.11}$ & 4.1$^{+1.6}_{-1.5}$ &0.63\\
2017-11-21T06:42:27 & 1110010108 & 3.0$^{+0.8}_{-0.8}$ & 3.0$^{+0.8}_{-0.6}$ & 0.56$^{+0.32}_{-0.28}$ & 57.79$^{+0.11}_{-0.10}$ & 4.0$^{+1.6}_{-1.5}$ &0.56\\
2017-12-12T06:09:20 & 1110010109 & 2.9$^{+1.7}_{-1.1}$ & 2.7$^{+1.0}_{-0.8}$ & $<$1.49 & 57.81$^{+0.22}_{-0.15}$ & 3.6$^{+1.0}_{-1.1}$ &0.73\\
2017-12-22T00:41:36 & 1110010110 & 2.7$^{+0.2}_{-0.2}$ & 3.5$^{+0.3}_{-0.3}$ & 0.53$^{+0.10}_{-0.09}$ & 57.83$^{+0.03}_{-0.03}$ & 5.4$^{+0.6}_{-0.6}$ &0.78\\
2017-12-30T00:19:52 & 1110010111 & 3.3$^{+0.8}_{-0.7}$ & 3.3$^{+0.9}_{-0.7}$ & 0.62$^{+0.32}_{-0.28}$ & 57.86$^{+0.11}_{-0.10}$ & 5.1$^{+1.3}_{-1.3}$ &0.60\\
2018-01-18T20:51:00 & 1110010114 & 3.2$^{+1.0}_{-1.0}$ & 3.1$^{+1.4}_{-0.7}$ & $<$0.42 & 57.88$^{+0.14}_{-0.15}$ & 4.6$^{+2.0}_{-1.7}$ &0.54\\
\hline
\end{tabular}
\end{center}
The columns in order describe the (1) \textit{NICER} observation date and time in UT, (2) unique observation number associated with the observation, (3) absorbing column density, (4) plasma energy (temperature), (5) Iron abundance relative to Solar, (6) log of the emission measure, (7) observed flux in the 2--10 keV band assuming a distance of 2.6 kpc, and (8) the reduced $\chi^{2}$ value associated with the best fit model.
Table locations with missing values indicate the spectral parameter or both confidence values hit a hard minimum or maximum when fitting and thus are not reliable. Spectral parameters missing a single upper or lower confidence value are reported as lower or upper limits, respectively.
X-ray spectral models and fitting procedure are described in more detail in section \ref{sec:obs}.
The full table is available in machine readable format.
\label{table:210}
\end{table*}

\begin{table*}[]
%\centering
\begin{center}
\begin{singlespace}
\caption{
Best-fit model parameters to the 5--10 keV \textit{NICER} X-ray spectra of Eta Car between July 2017 and July 2020 and their 90\% confidence values.}
\end{singlespace}
\begin{tabular}{c c c c c c c c c}
\hline\hline
Date & Obs.\,ID. & $N_\mathrm{H}$ & $kT$  & Fe & log(EM) & Flux $\times10^{-11}$ & Int. Flux $\times10^{-11}$ & $\chi^2_{\rm red}$\\
$[$UT$]$ &  & $[$$10^{22}$ cm$^{-2}]$& $[$keV$]$  & $[$Z$_{\odot}]$ & $[$cm$^{-3}]$ & $[$erg s$^{-1}$ cm$^{-2}]$ & $[$erg s$^{-1}$ cm$^{-2}]$ & \\
 \hline
2017-07-20T02:59:20 & 1110010101 & $<$64 & 1.8$^{+3.3}_{-0.8}$ & 0.69$^{+1.52}_{-0.41}$ & 58.28$^{+1.04}_{-0.77}$ & 1.2$^{+4.1}_{-1.1}$ &2.1$^{+6.7}_{-1.9}$ &0.53\\
2017-07-21T00:36:20 & 1110010102 &  --- & 4.9$^{+1.9}_{-2.7}$ &  --- & 57.53$^{+0.84}_{-0.11}$ & 1.5$^{+1.2}_{-0.9}$ &2.0$^{+1.3}_{-1.1}$ &0.60\\
2017-07-22T04:23:20 & 1110010103 & 59$^{+38}_{-39}$ & 1.3$^{+0.6}_{-0.4}$ & 0.41$^{+0.48}_{-0.22}$ & 59.29$^{+0.84}_{-0.85}$ & 2.2$^{+7.2}_{-1.8}$ &7.2$^{+19}_{-6.2}$ &0.65\\
2017-07-24T10:25:00 & 1110010105 &  --- & 2.3$^{+4.1}_{-1.4}$ & 0.60$^{+2.13}_{-0.45}$ & 58.09$^{+1.91}_{-0.58}$ & 0.9$^{+3.7}_{-0.8}$ &2.4$^{+6.4}_{-2.2}$ &0.67\\
2017-10-05T02:11:20 & 1110010106 & $<$106 & 1.2$^{+1.8}_{-0.4}$ & 0.37$^{+0.77}_{-0.26}$ & 59.32$^{+0.93}_{-1.54}$ & 2.2$^{+10}_{-2.0}$ &7.8$^{+27}_{-6.9}$ &0.63\\
2017-11-21T06:42:27 & 1110010108 &  --- & 1.9$^{+3.0}_{-0.8}$ & 0.73$^{+1.10}_{-0.44}$ & 58.06$^{+0.99}_{-0.64}$ & 0.9$^{+3.2}_{-0.8}$ &1.4$^{+4.4}_{-1.2}$ &0.57\\
2017-12-12T06:09:20 & 1110010109 & $<$95 & 1.3$^{+1.0}_{-0.7}$ &  --- & 58.46$^{+1.61}_{-0.64}$ & 0.5$^{+2.3}_{-0.5}$ &1.5$^{+4.8}_{-1.3}$ &1.74\\
2017-12-22T00:41:36 & 1110010110 & 29$^{+17}_{-19}$ & 2.2$^{+0.9}_{-0.5}$ & 0.38$^{+0.14}_{-0.10}$ & 58.44$^{+0.39}_{-0.46}$ & 2.0$^{+3.4}_{-1.5}$ &3.6$^{+5.1}_{-2.8}$ &0.58\\
2017-12-30T00:19:52 & 1110010111 & $<$63 & 1.7$^{+1.5}_{-0.6}$ & 0.77$^{+0.87}_{-0.46}$ & 58.34$^{+0.96}_{-0.55}$ & 1.0$^{+2.9}_{-0.9}$ &1.8$^{+3.8}_{-1.4}$ &0.72\\
2018-01-18T20:51:00 & 1110010114 & $<$99 & 1.6$^{+5.5}_{-1.0}$ & $<$0.80 & 58.76$^{+1.45}_{-1.28}$ & 1.5$^{+8.3}_{-1.3}$ &4.7$^{+15}_{-3.9}$ &0.37\\
\hline
\end{tabular}
\end{center}
The columns in order describe the (1) \textit{NICER} observation date and time in UT, (2) unique observation number associated with the observation, (3) absorbing column density, (4) plasma energy (temperature), (5) Iron abundance relative to Solar, (6) log of the emission measure, (7) observed flux in the 5--10 keV band assuming a distance of 2.6 kpc, (8) the intrinsic (i.e., unabsorbed) flux in the 5--10 keV band, and (9) the reduced $\chi^{2}$ value associated with the best fit model. 
Table locations with missing values indicate the spectral parameter or both confidence values hit a hard minimum or maximum when fitting and thus are not reliable. Spectral parameters missing a single upper or lower confidence value are reported as lower or upper limits, respectively.
X-ray spectral models and fitting procedure are described in more detail in section \ref{sec:obs}
The full table is available in machine readable format.
\label{table:510}
\end{table*}

%%%%%%%%%%%%%%%%%%%%%%%%%%%%%%%%%%%%%%%%%%

Although the abundances of the X-ray emitting plasma associated with $\eta$ Car are uncertain (and are likely enriched in He and N; e.g., \citealt{Gulletal2020}), we note that our assumption of solar metal abundances (apart from Fe) is unlikely to affect these fit results. This is mainly because the 5--10~keV region is largely devoid of strong metal emission lines, apart from the 6.4~keV and 7.1~keV Fe lines. Indeed, based on tests performed on a subset of observations, we confirmed that the results for X-ray flux and $N_{\rm H}$ are insensitive to the assumed metallicity.

\begin{figure*}
\includegraphics[trim= 1.3cm 0.0cm 2.0cm 1.5cm,clip=true,width=0.5\textwidth]{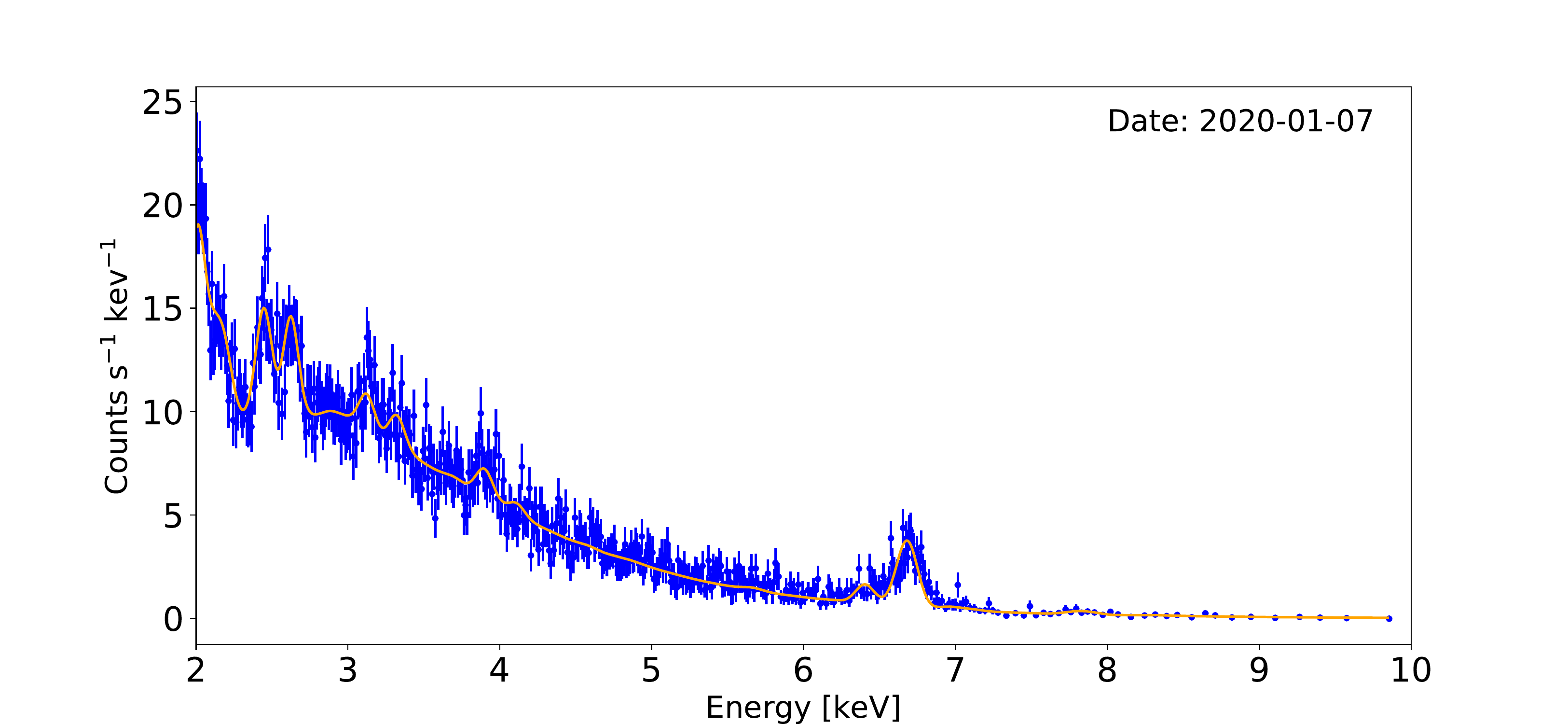} %l b r t
\includegraphics[trim= 1.3cm 0.0cm 2.0cm 1.5cm,clip=true,width=0.5\textwidth]{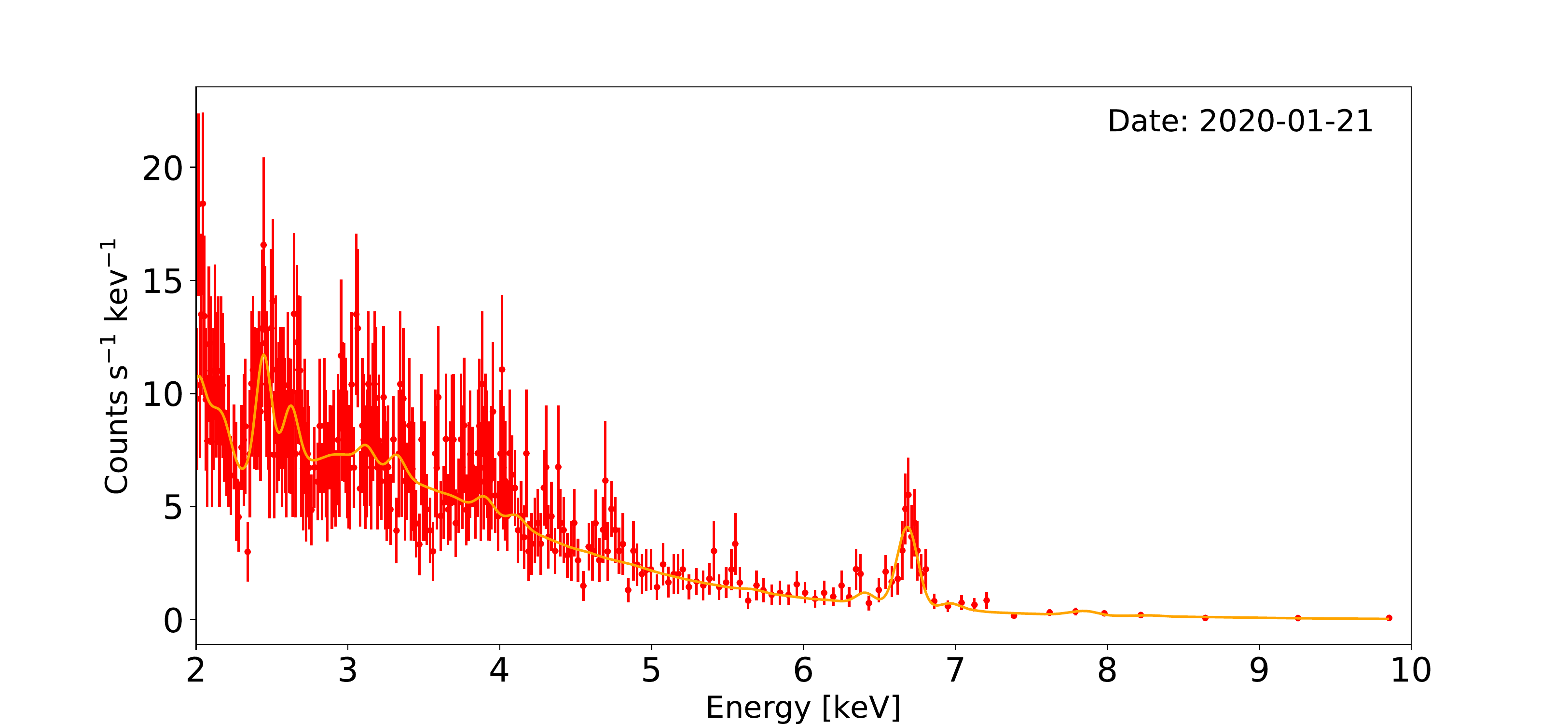}
\includegraphics[trim= 1.3cm 0.0cm 2.0cm 1.5cm,clip=true,width=0.5\textwidth]{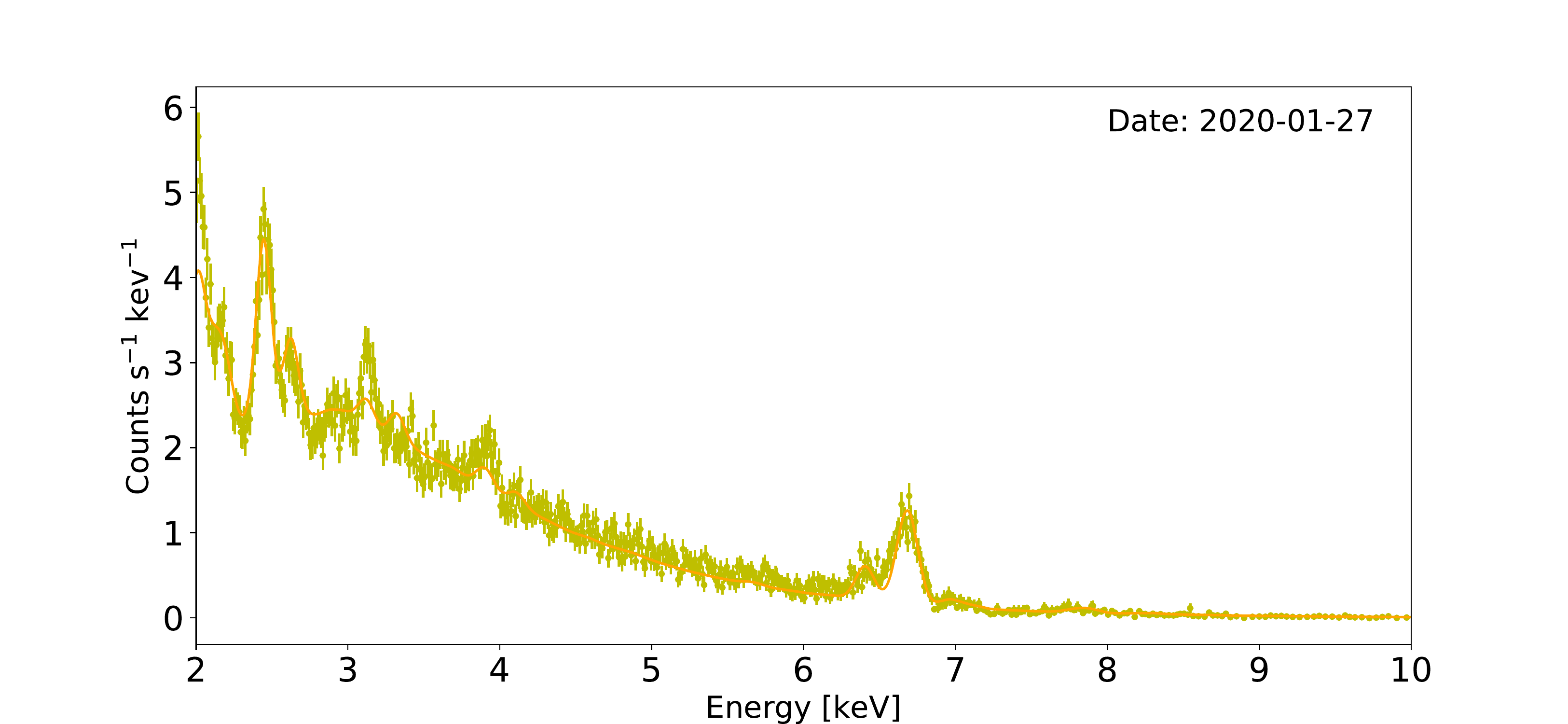}
\includegraphics[trim= 1.3cm 0.0cm 2.0cm 1.5cm,clip=true,width=0.5\textwidth]{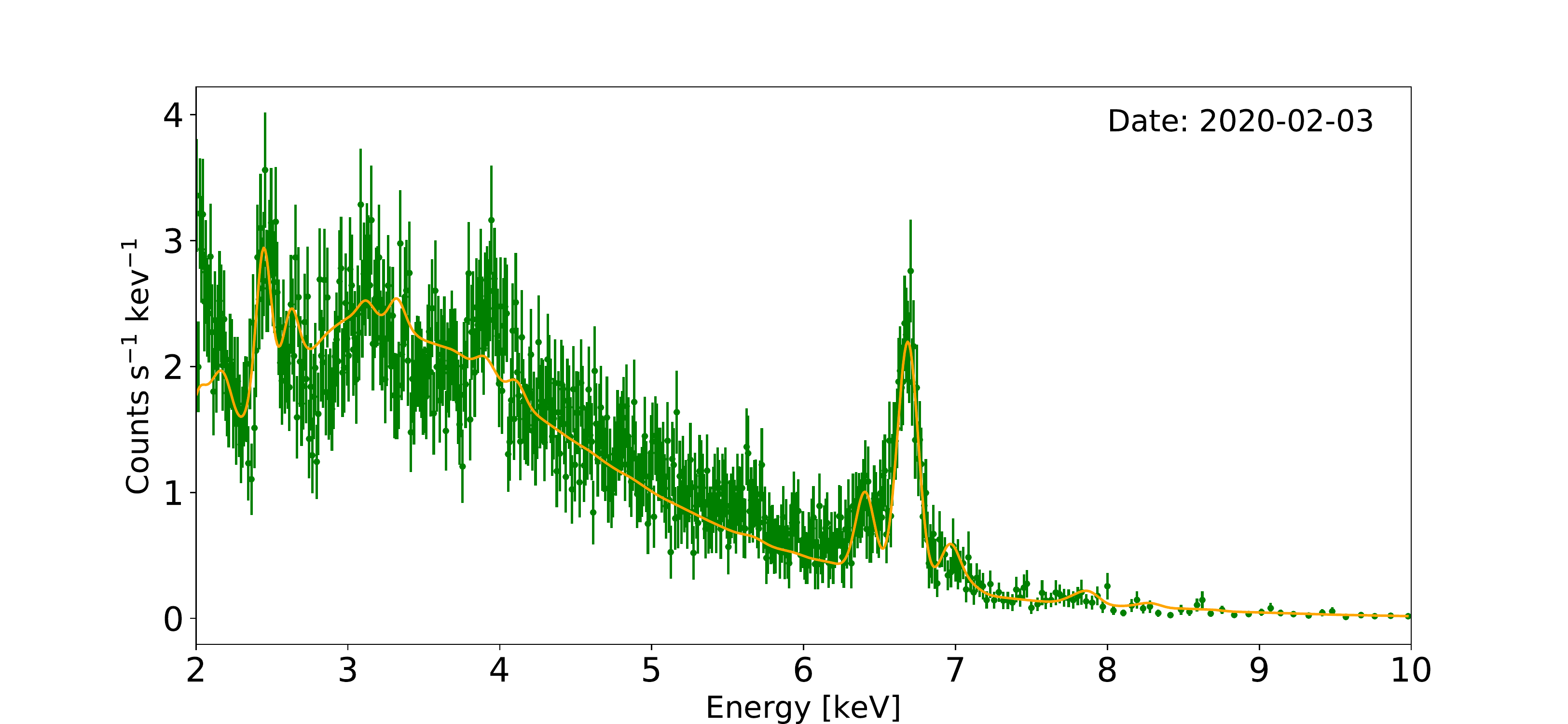}
\includegraphics[trim= 1.3cm 0.0cm 2.0cm 1.5cm,clip=true,width=0.5\textwidth]{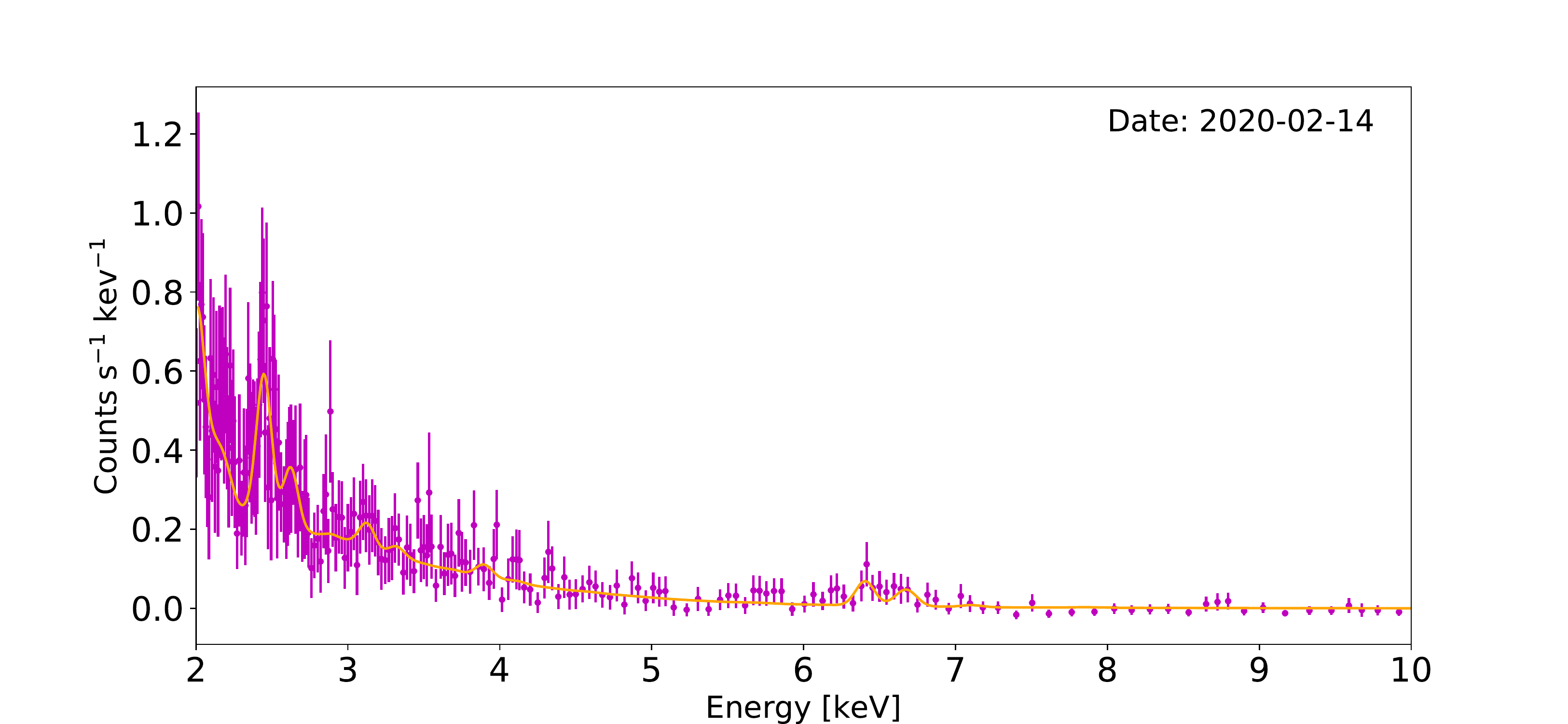}
\includegraphics[trim= 1.3cm 0.0cm 2.0cm 1.5cm,clip=true,width=0.5\textwidth]{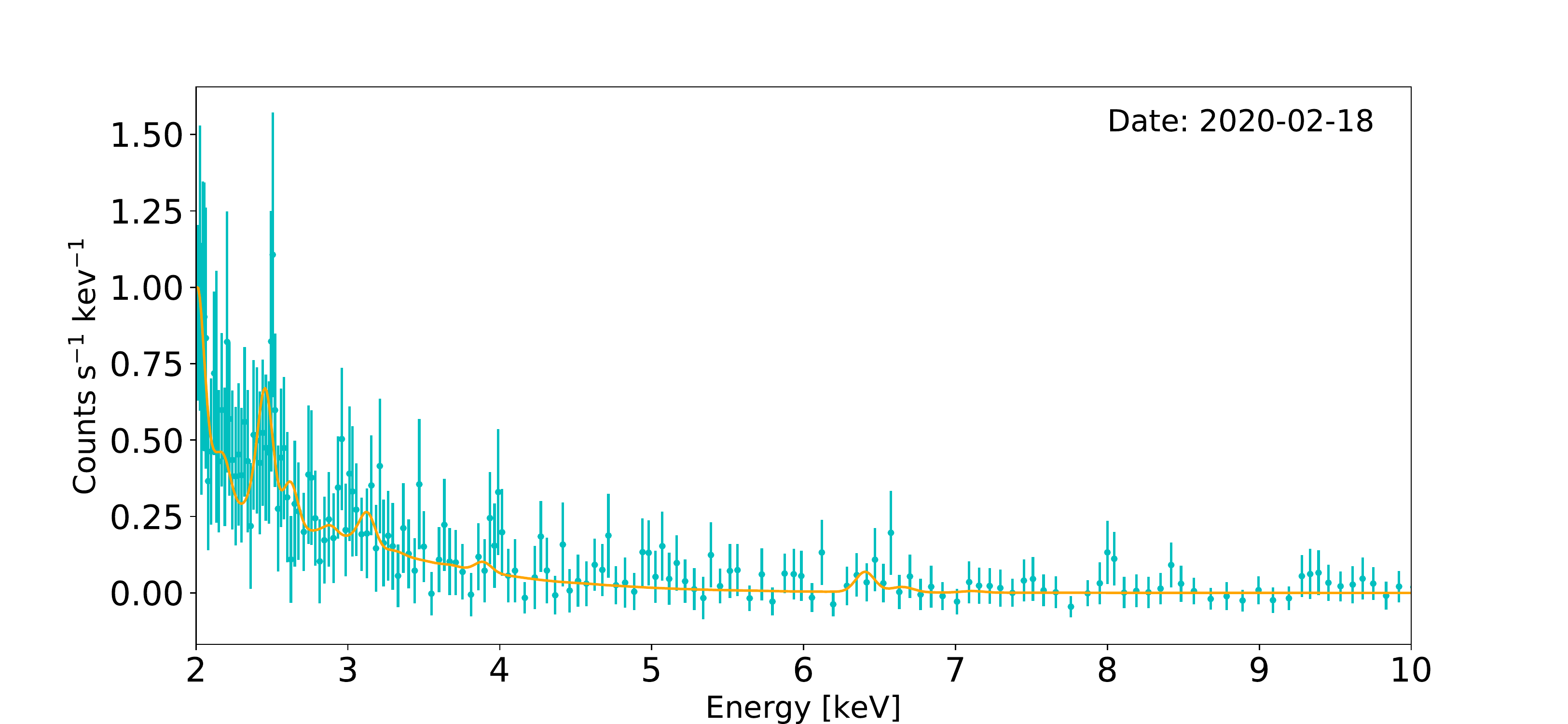}
\begin{singlespace}
\caption{X-ray spectral fits to the 2--10 keV \textit{NICER} data of $\eta$ Car across the 2020 X-ray minimum. Spectra are grouped to require a minimum SNR$=$3 where dates and colors correspond to the same format as Figure \ref{fig:xrayspectra}. The best-fit models are shown in orange and described in section \ref{sec:obs}.}
\end{singlespace}
\label{fig:xrayspectra210}
\end{figure*}

\begin{figure*}
\includegraphics[trim= 1.3cm 0.0cm 2.0cm 1.5cm,clip=true,width=0.5\textwidth]{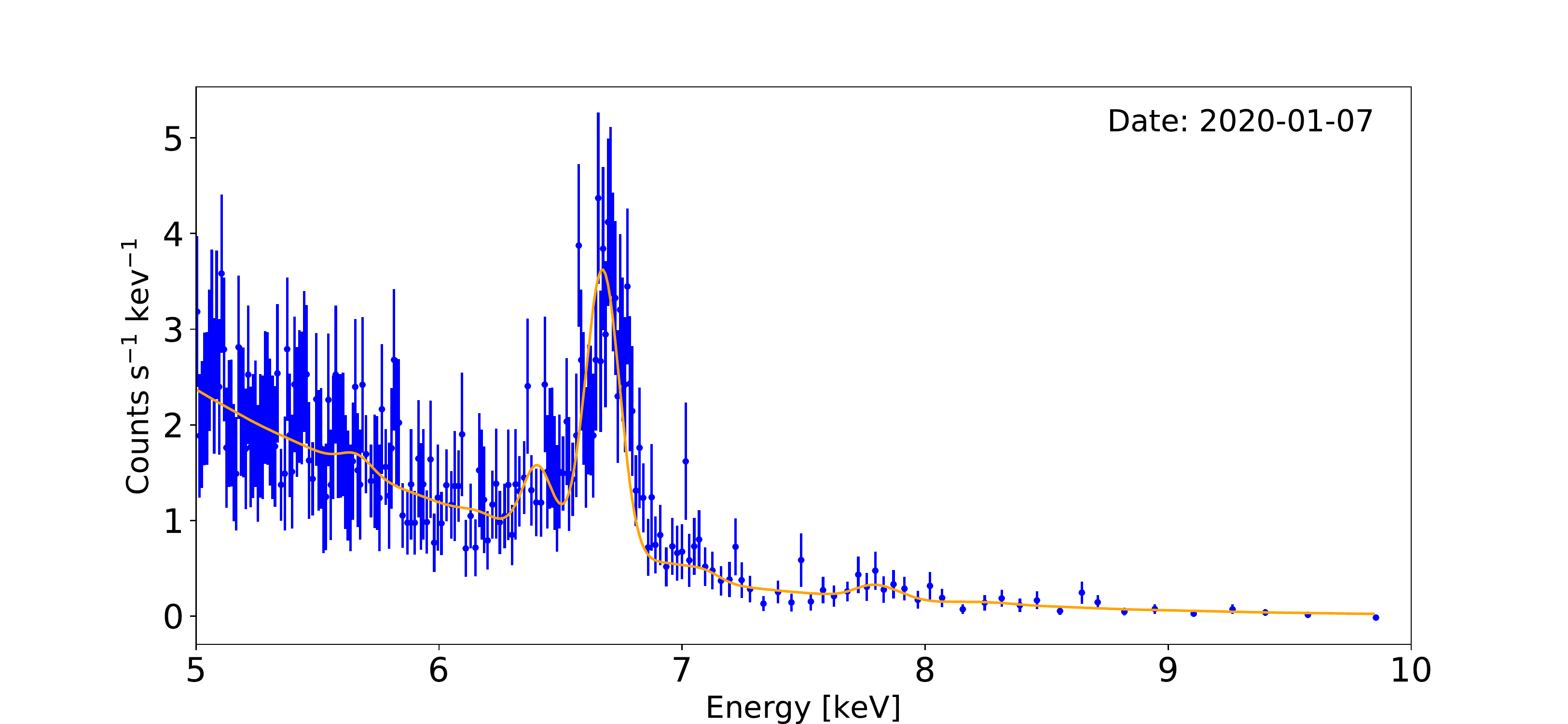} %l b r t
\includegraphics[trim= 1.3cm 0.0cm 2.0cm 1.5cm,clip=true,width=0.5\textwidth]{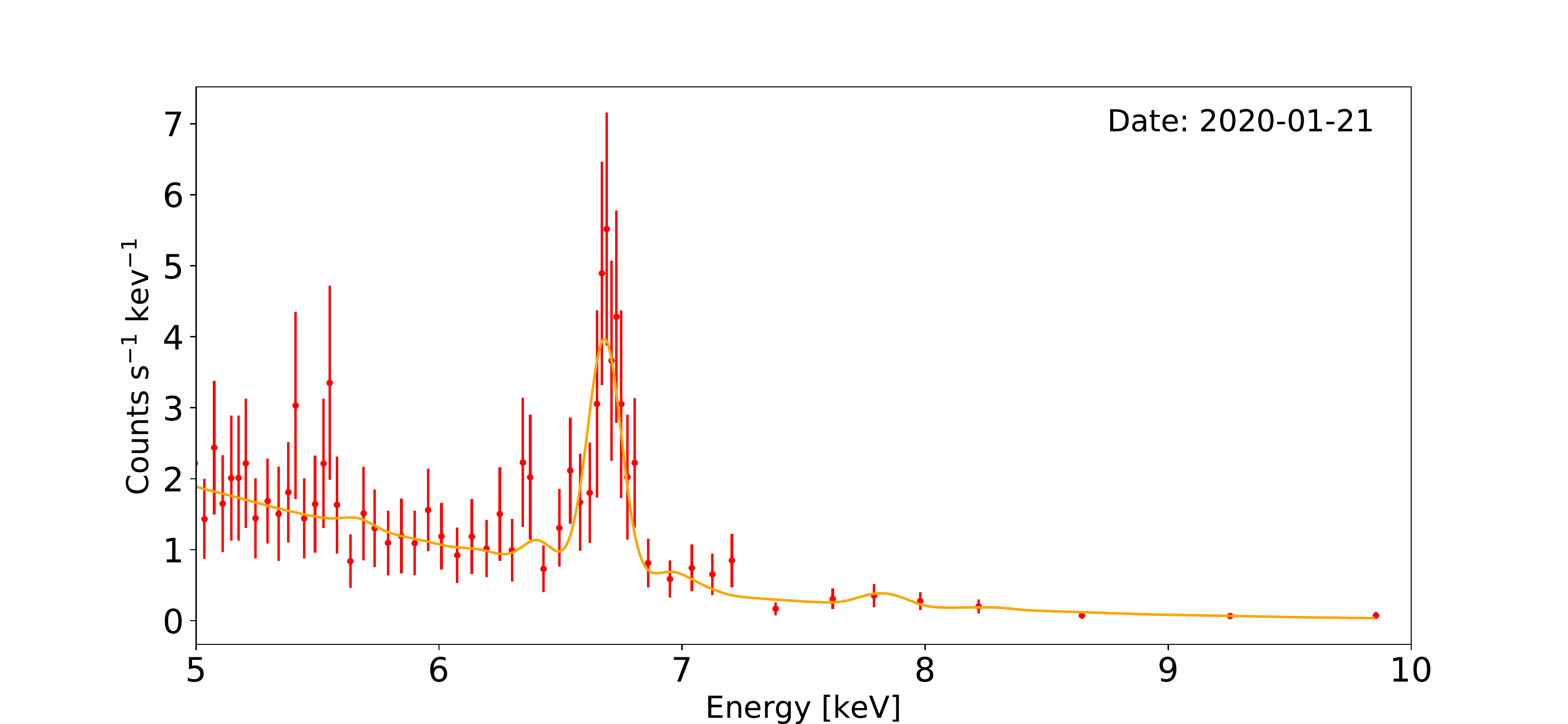}
\includegraphics[trim= 1.3cm 0.0cm 2.0cm 1.5cm,clip=true,width=0.5\textwidth]{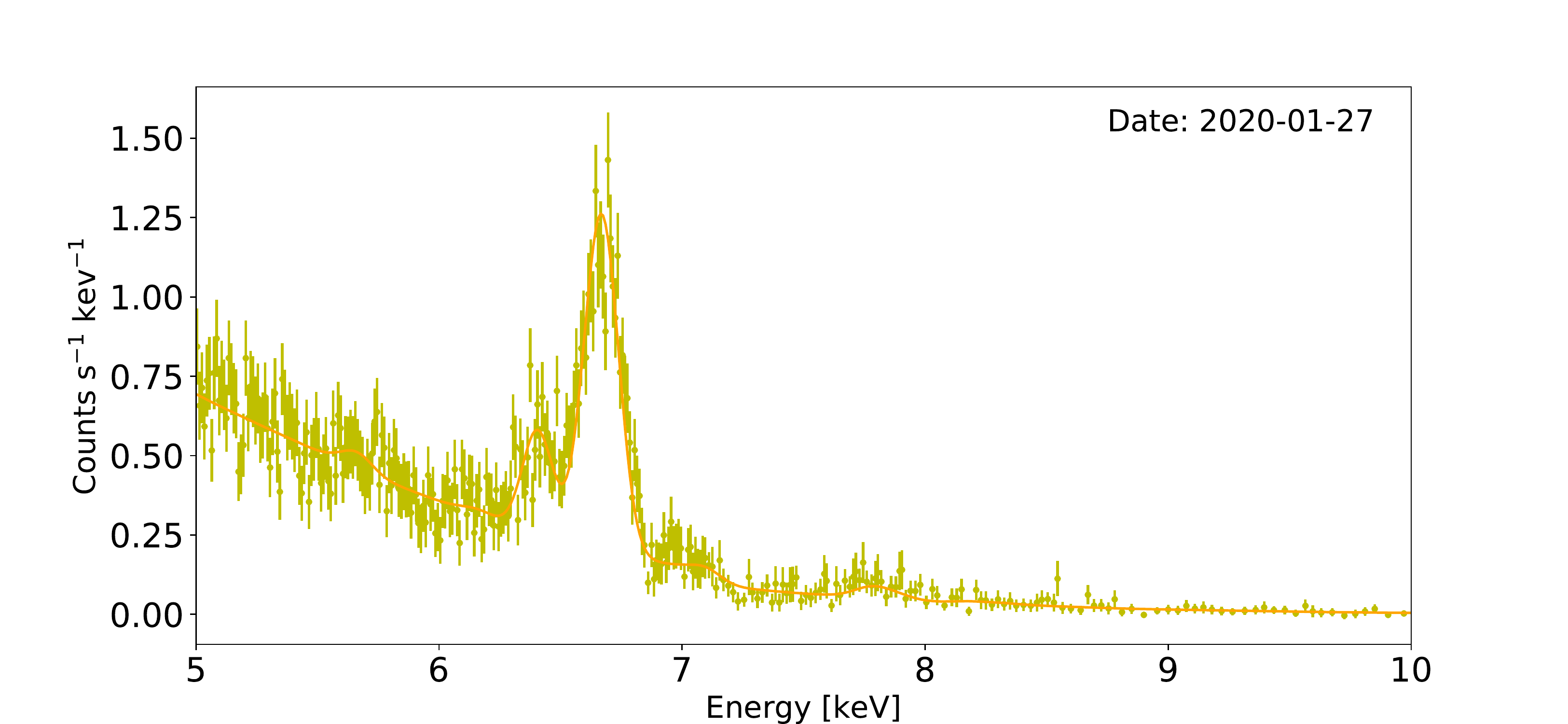}
\includegraphics[trim= 1.3cm 0.0cm 2.0cm 1.5cm,clip=true,width=0.5\textwidth]{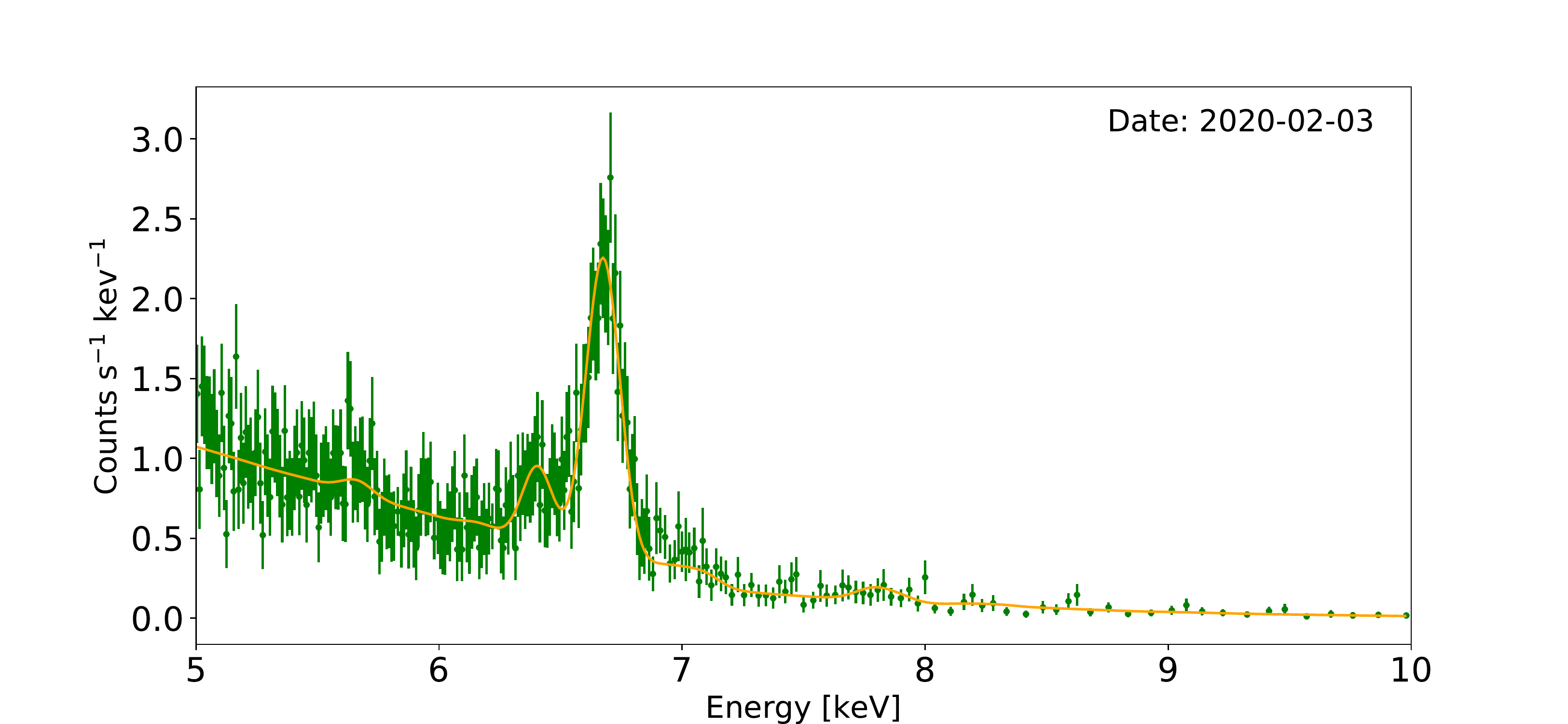}
\includegraphics[trim= 1.3cm 0.0cm 2.0cm 1.5cm,clip=true,width=0.5\textwidth]{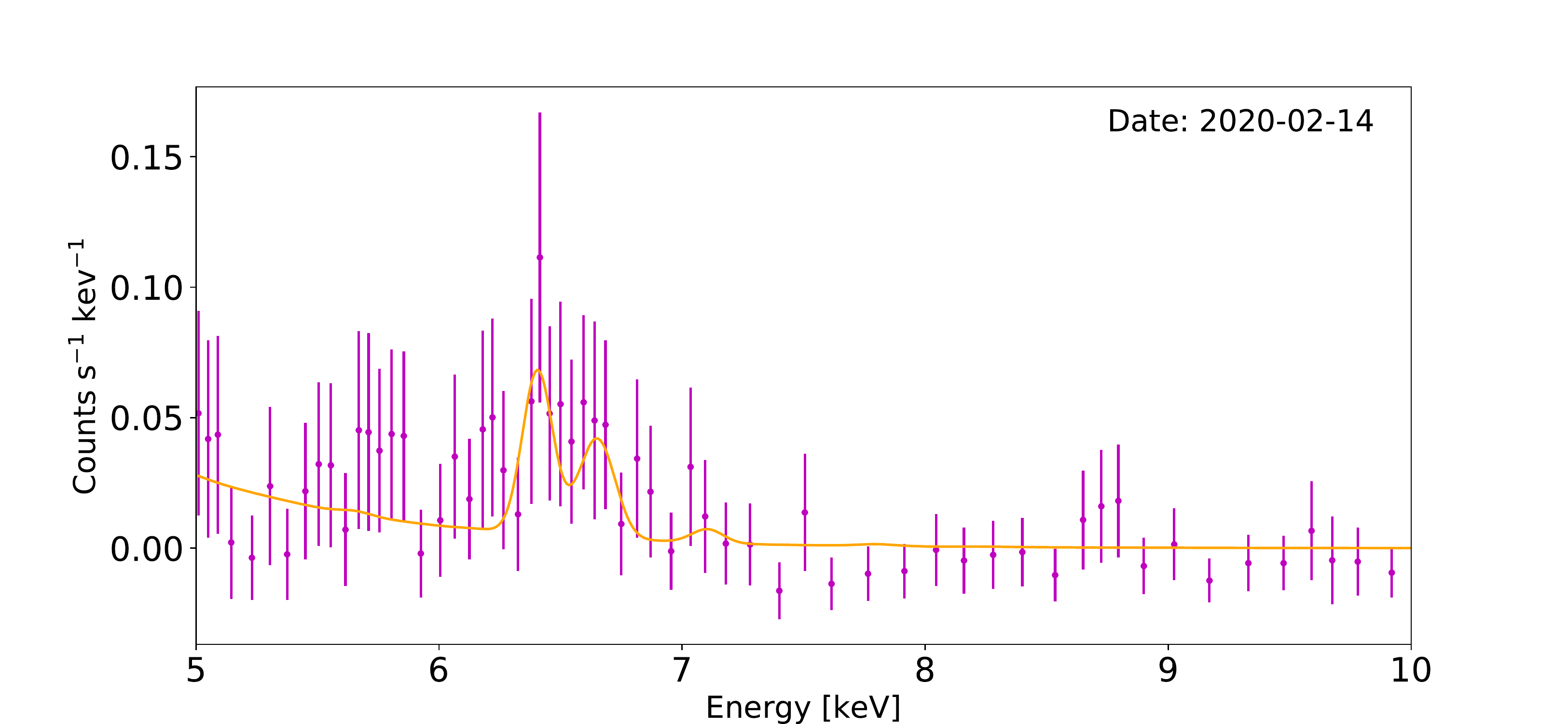}
\includegraphics[trim= 1.3cm 0.0cm 2.0cm 1.5cm,clip=true,width=0.5\textwidth]{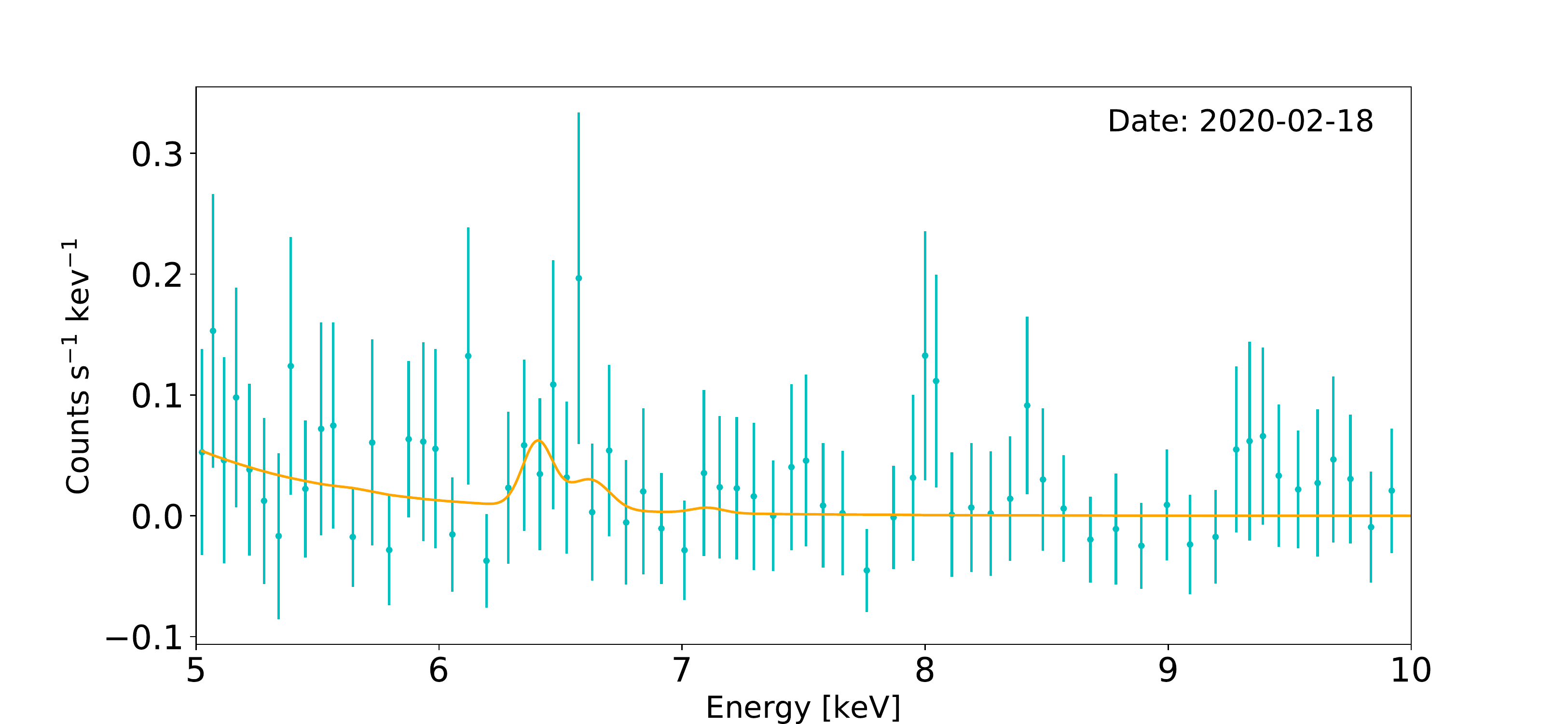}
\begin{singlespace}
\caption{X-ray spectral fits to the 5--10 keV \textit{NICER} data of $\eta$ Car across the 2020 X-ray minimum. Spectra are grouped to require a minimum SNR$=$3 where dates and colors correspond to the same format as Figure \ref{fig:xrayspectra}. The best-fit models are shown in orange and described in section \ref{sec:obs}.}
\end{singlespace}
\label{fig:xrayspectra510}
\end{figure*}

Figure \ref{fig:xrayspectra} shows selected $\eta$~Car \textit{NICER} spectra for a number of notable epochs.
January 7, 21, 27, and February 3 (2020) are times of strong peaks in the lightcurve, which are usually referred to as flares.
February 14 and 18 are representative specta taken during the X-ray minimum.
Figs. \ref{fig:xrayspectra210} and \ref{fig:xrayspectra510} show our fit to the \textit{NICER} 2--10 keV and 5--10 keV energy ranges, respectively.
%
%FFFFFFFFFFFFFFFFFFFFFFFFFFFFFFFFFFFFFFFFFFFFFFFFFFFFFFFFFFFFFFFFFFF
%FFFFFFFFFFFFFFFFFFFFFFFFFFFFFFFFFFFFFFFFFFFFFFFFFFFFFFFFFFFFFFFFFFF
%
%FFFFFFFFFFFFFFFFFFFFFFFFFFFFFFFFFFFFFFFFFFFFFFFFFFFFFFFFFFFFFFFFFFF
%FFFFFFFFFFFFFFFFFFFFFFFFFFFFFFFFFFFFFFFFFFFFFFFFFFFFFFFFFFFFFFFFFFF
%
%FFFFFFFFFFFFFFFFFFFFFFFFFFFFFFFFFFFFFFFFFFFFFFFFFFFFFFFFFFFFFFFFFFF
%FFFFFFFFFFFFFFFFFFFFFFFFFFFFFFFFFFFFFFFFFFFFFFFFFFFFFFFFFFFFFFFFFFF

\begin{figure*}
\includegraphics[trim= 0.0cm 0.0cm 0.0cm 0.0cm,clip=true,width=0.99\textwidth]{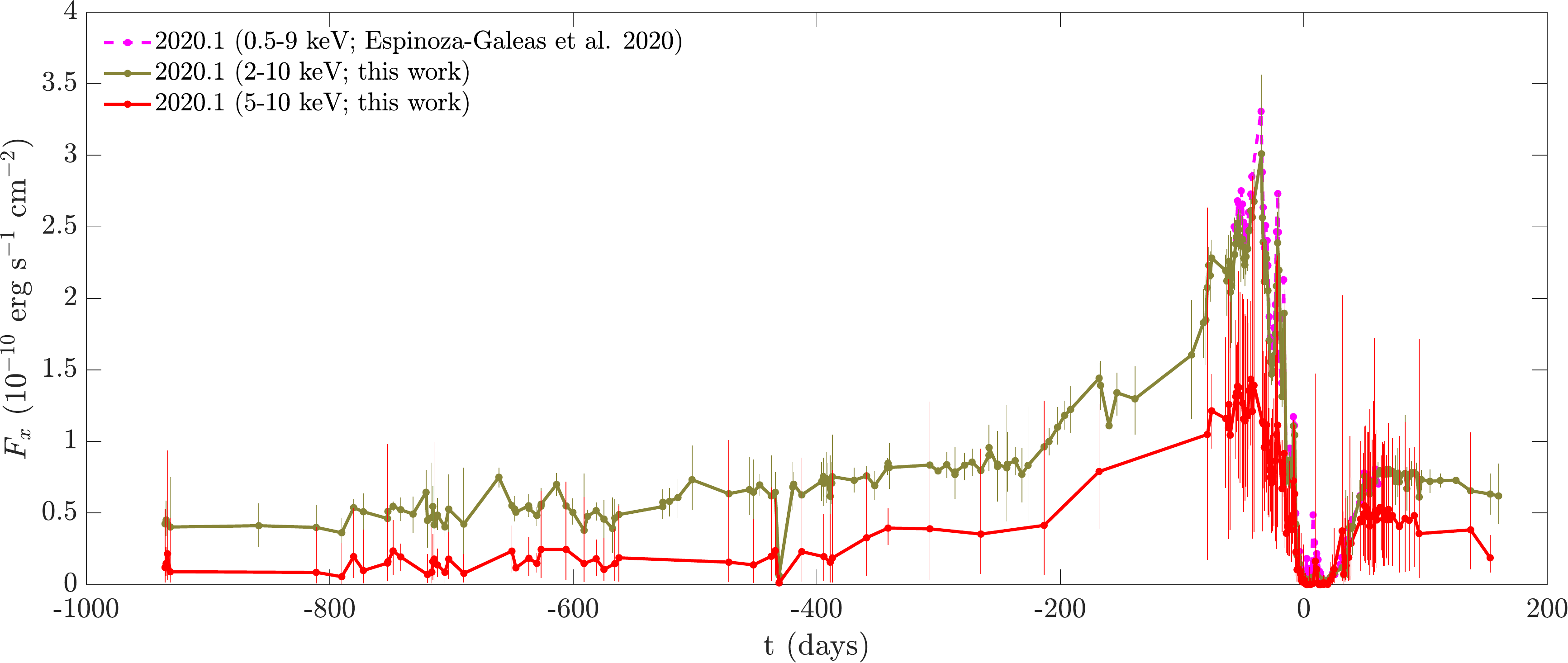}  %l b r t
\includegraphics[trim= 0.0cm 0.0cm 0.0cm 0.0cm,clip=true,width=0.99\textwidth]{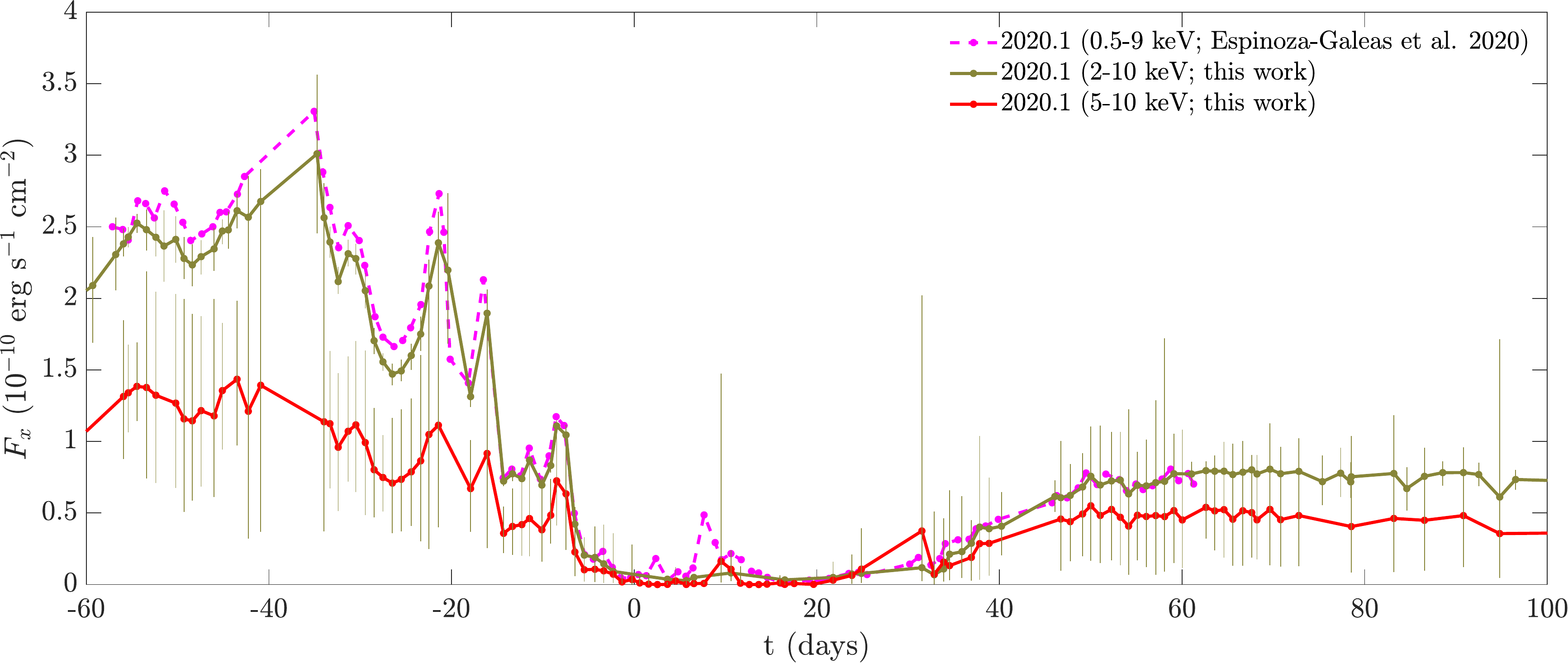}  %l b r t
\begin{singlespace}
\caption{
{Upper panel:} The \textit{NICER} 2-10 keV and 5--10 keV (hard X-ray) light curve of $\eta$~Car and 90\% confidence values for both ranges.
\textit{Lower panel:} zoom in on the 2020 X-ray minimum.
We compare our fitted flux to the results of \cite{Espinoza-Galeasetal2020}. Note that error bars were not given for the data in \cite{Espinoza-Galeasetal2020}.
The time axis is set to the beginning of periastron passage; $t=0$ is on Feb 10, 2020 (JD$=2458890$).
We see that the X-ray minimum flare obtained by \cite{Espinoza-Galeasetal2020} on $t=8 \days$ is not present in our results.
The figure also demonstrates that the for for the 5--10 keV energy range results in better error estimate than for the 2--10 keV energy range.
}
\end{singlespace}
\label{fig:xray}
\end{figure*}

Figure \ref{fig:xray} presents the resulting broad-band (2--10 keV) and hard (5--10 keV) X-ray flux variations (hereafter we will refer to them simply as light curves) during the 2020 spectroscopic X-ray minimum.

We present only observations that yielded reduced $\chi^{2}$ values in the range $0.4\leq\chi^2_{\rm red}\leq2.5$, and $N_{\rm H}$ where 90\% confidence intervals were determined, as well as those observations with very low flux $F\leq 10^{-11} \rm{erg~s^{-1}~cm^{-2}}$ even if the aforementioned conditions for $\chi^2_{\rm red}$ and $N_{\rm H}$ were not fulfilled.
Adopting a distance of $2.6 \kpc$ \citep{Davidsonetal2018a}, flux of $10^{-10} ~\rm{erg~s^{-1}~cm^{-2}}$ corresponds to luminosity of $8.09\times10^{34} ~\rm{erg~s^{-1}}$.
Figure \ref{fig:xray} also shows the 0.5--10 keV X-ray light curve reported by \cite{Espinoza-Galeasetal2020}, where we have combined the two sets of observations.
In the \cite{Espinoza-Galeasetal2020} light curve, the X-ray minimum appears to be interrupted by a flare, with a peak flux $3.3\times10^{-11} ~\rm{erg~s^{-1}~cm^{-2}}$ and a duration of $\simeq 5 \days$.
However, in our reanalysis of the 2--10 keV \textit{NICER} data, we did not recover a flare during this time period (compare the solid green and dashed magenta curves in Figure~ \ref{fig:xray}).
The S/N ratios in the spectra during X-ray minimum are poor, such that small differences in background subtraction methods may account for the differences between the respective light curves.
\cite{Espinoza-Galeasetal2020} ATel do not detail background subtraction methods and thus it is difficult to identify the discrepancy in the flare detection. As described in section \ref{sec:obs}, we use the most up to date calibration files as well as a suggested background determination method from the \textit{NICER} team (Remillard et al, in prep\footnote{See \textit{NICER} background estimator tools page: \url{https://heasarc.gsfc.nasa.gov/docs/nicer/tools/nicer_bkg_est_tools.html}}).
We note that the apparent flare reported by \cite{Espinoza-Galeasetal2020} cannot originate from the 0.5--2 keV spectral component, since this component arises from $\eta$~Car's surrounding, extended emission and hence does not vary during the X-ray minimum \citep{Hamaguchietal2007}.
%
%FFFFFFFFFFFFFFFFFFFFFFFFFFFFFFFFFFFFFFFFFFFFFFFFFFFFFFFFFFFFFFFFFFF

%FFFFFFFFFFFFFFFFFFFFFFFFFFFFFFFFFFFFFFFFFFFFFFFFFFFFFFFFFFFFFFFFFFF

\begin{figure*}
\includegraphics[trim= 0.0cm 0.0cm 0.0cm 0.0cm,clip=true,width=0.99\textwidth]{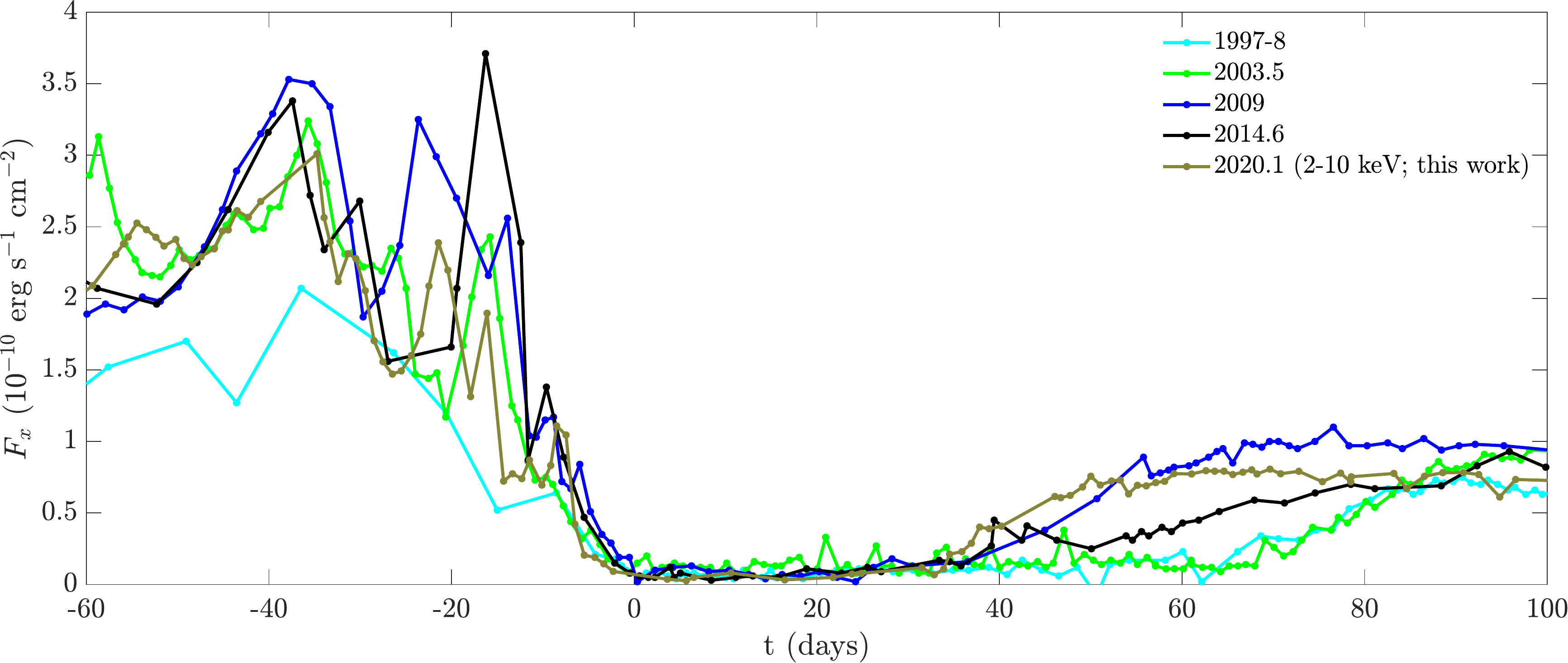}  %l b r t
\begin{singlespace}
\caption{\
Comparison of the X-ray minimum for the last 5 X-ray minima of $\eta$~Car. 
Data for earlier cycles is adopted from \cite{Corcoranetal2017}.
The time scale is fixed and the beginning of the {1997--8} X-ray minimum and a period of 2023 days is used to fold the data.
The duration of the 2020 X-ray minimum can be considered to be anything in the range $\simeq25$--$37 \days$, depending on the definition.
The recovery from the 2020 X-ray minimum occurs at the steepest slope.
}
\end{singlespace}
\label{fig:xray_cycles}
\end{figure*}

The duration of the X-ray minimum has varied in the last 4 cycles where it was closely monitored \citep{Corcoranetal2017}.
Figure \ref{fig:xray_cycles} shows a comparison of the last 5 cycles, focused on the X-ray minimum.
As can be seen in Figure \ref{fig:xray}, the duration of the X-ray minimum in the 2020 cycle is not well constrained as the recovery is gradual and has some fluctuations. The duration can be considered to be anything in the range $\simeq25$--$37 \days$.
The hard component has a clearer recovery, and its duration is $23 \days$.

The recovery from the 2020 X-ray minimum occurs at the steepest slope amongst the 5 cycles.
A quantifying criterion that takes the slope of the recovery into account would indicate that the present cycle had the shortest minimum.
If, for example, we consider the duration of the X-ray minimum to be the time where the flux is $F \leq 5 \times 10^{-11} \rm{erg~s^{-1}~cm^{-2}}$ then the 2020 X-ray minimum is the shortest.
Hereafter we will refer to this finding about the 2020 Minimum simply as `fastest recovery'.
%
%FFFFFFFFFFFFFFFFFFFFFFFFFFFFFFFFFFFFFFFFFFFFFFFFFFFFFFFFFFFFFFFFFFF
%FFFFFFFFFFFFFFFFFFFFFFFFFFFFFFFFFFFFFFFFFFFFFFFFFFFFFFFFFFFFFFFFFFF

\begin{figure*}
\includegraphics[trim= 0.0cm 0.0cm 0.0cm 0.0cm,clip=true,width=0.95\textwidth]{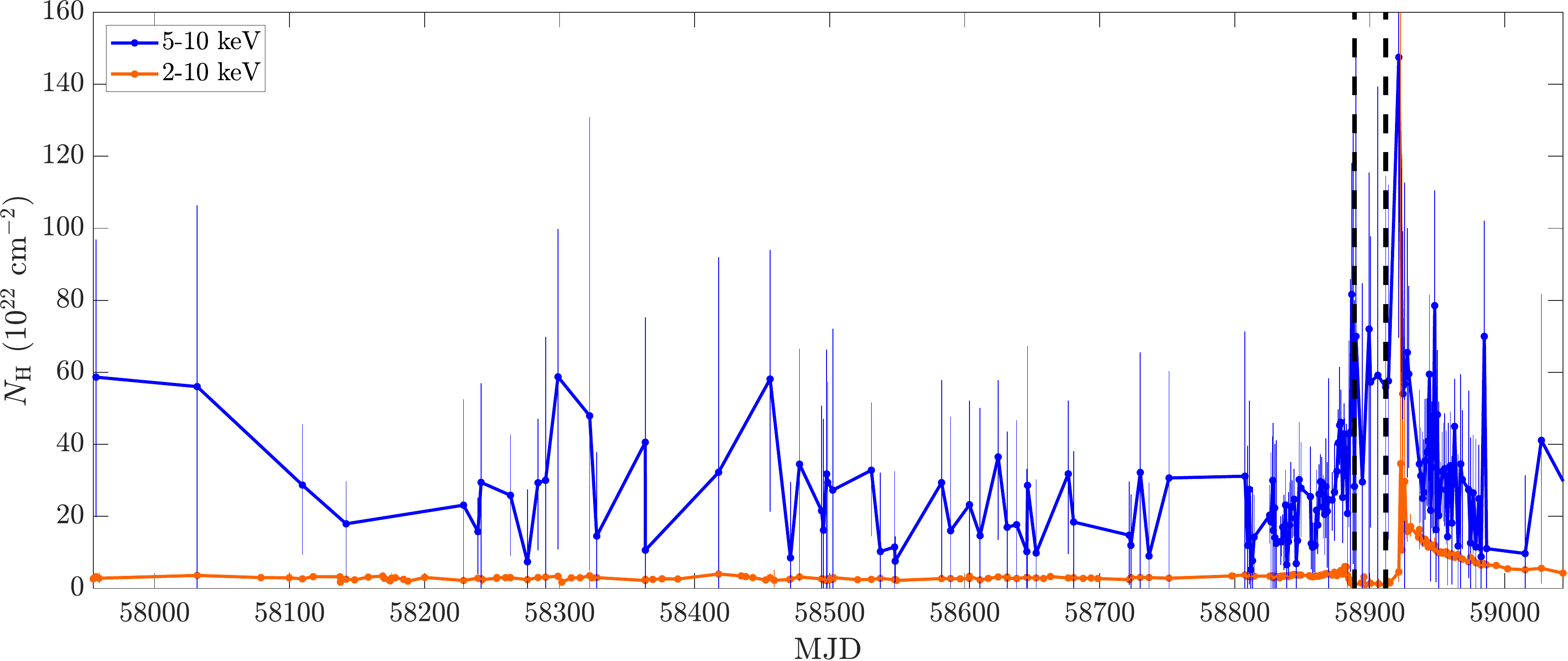}  %l b r t
\includegraphics[trim= 0.0cm 0.0cm 0.0cm 0.0cm,clip=true,width=0.97\textwidth]{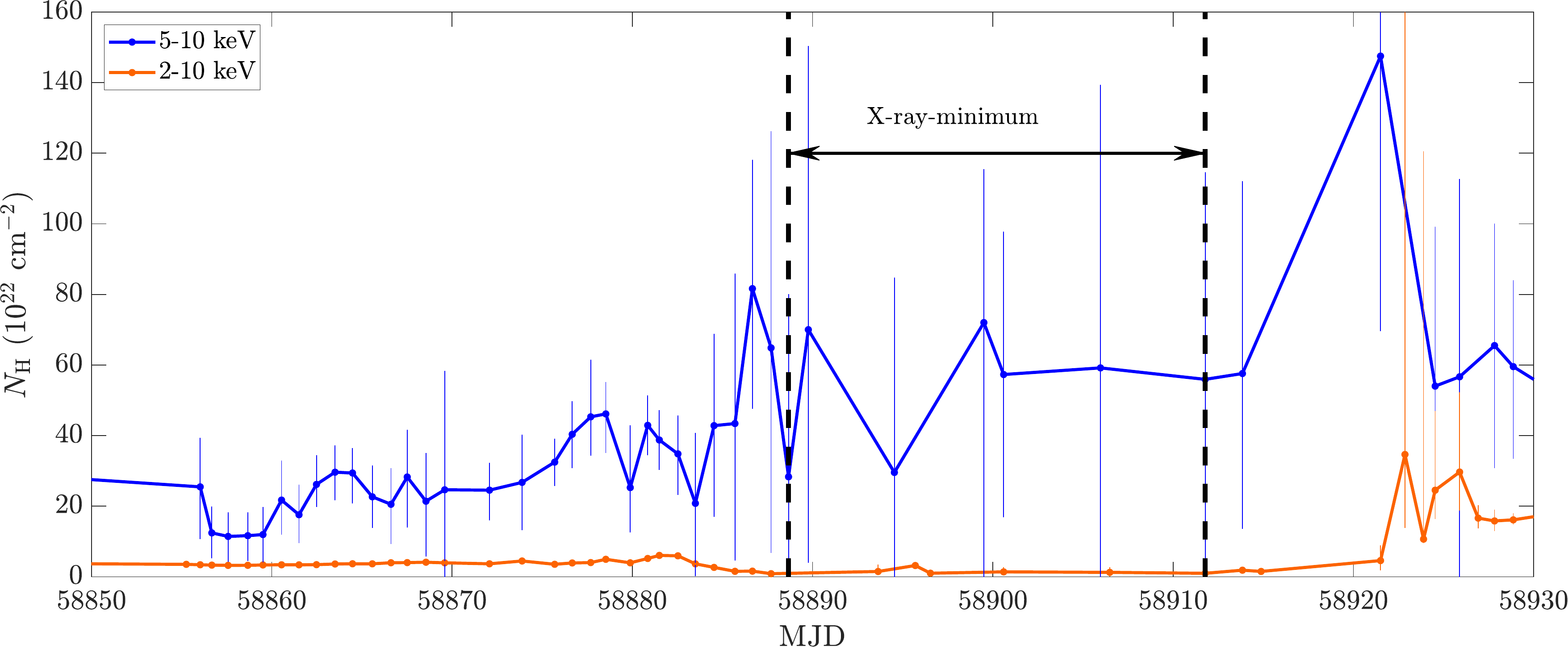}
\begin{singlespace}
\caption{\textit{Upper panel:} The derived hydrogen column density $N_{\rm H}$ for the 2--10 keV and 5--10 keV energy ranges, and 90\% confidence values bot both ranges.
The time axis begins about two years before periastron passage, where the stars are very far from each other on their $P=2023$ days orbit.
Dashed vertical lines in both panels represent the beginning and end of the X-ray minimum (note the uncertainties in the exit date that we discuss in the text).
\textit{Lower panel:} zoom in on the 2020 X-ray minimum (close to periastron passage).
}
\end{singlespace}
\label{fig:N_H_210510}
\end{figure*}

\begin{figure*}
\begin{center}
\includegraphics[trim= 0.0cm 0.0cm 0.0cm 0.0cm,clip=true,width=0.89\textwidth]{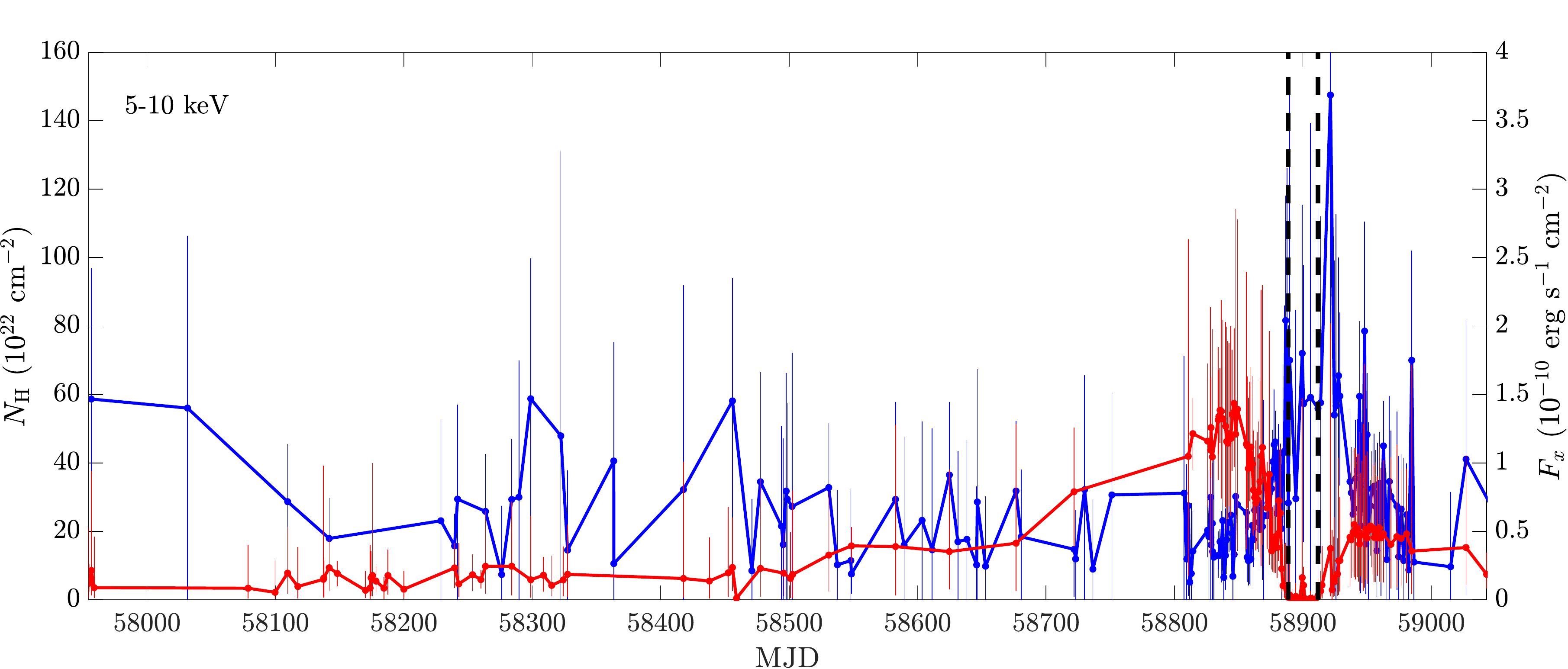}  %l b r t
\includegraphics[trim= 1.1cm 0.0cm 0.0cm 0.0cm,clip=true,width=1.00\textwidth]{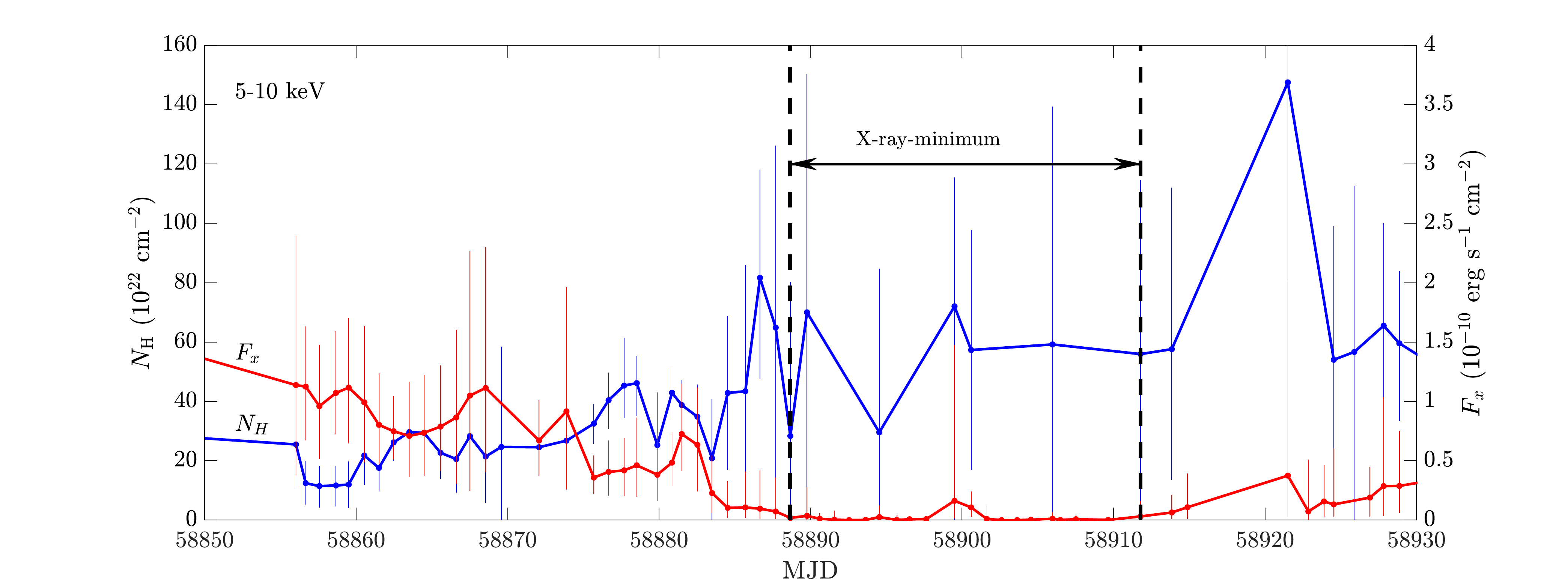}
\end{center}
\begin{singlespace}
\caption{\textit{Upper panel:} The derived hydrogen column density $N_{\rm H}$ of the hard X-ray $\eta$~Car from \textit{NICER} observations (left axis, blue line).
The time axis begins about two years before periastron passage, where the stars are very far from each other on their $P=2023$ days orbit.
Dashed vertical lines in both panels represent the beginning and end of the X-ray minimum.
\textit{Lower panel:} zoom in on the 2020 X-ray minimum (close to periastron passage).
The flux from Figure \ref{fig:xray} is also plotted in both panels (right axis, red line).
}
\end{singlespace}
\label{fig:N_H}
\end{figure*}

Figure \ref{fig:N_H_210510} shows the hydrogen column density $N_{\rm H}$ from the fitted \textit{NICER} spectra for the broad-band and hard X-ray ranges, while Figure \ref{fig:N_H} shows the hard X-ray flux and $N_{\rm H}$ together.
The duration of X-ray minimum for the hard X-ray component is marked in these two figures.
%
%FFFFFFFFFFFFFFFFFFFFFFFFFFFFFFFFFFFFFFFFFFFFFFFFFFFFFFFFFFFFFFFFFFF
%FFFFFFFFFFFFFFFFFFFFFFFFFFFFFFFFFFFFFFFFFFFFFFFFFFFFFFFFFFFFFFFFFFF
%
%FFFFFFFFFFFFFFFFFFFFFFFFFFFFFFFFFFFFFFFFFFFFFFFFFFFFFFFFFFFFFFFFFFF
%FFFFFFFFFFFFFFFFFFFFFFFFFFFFFFFFFFFFFFFFFFFFFFFFFFFFFFFFFFFFFFFFFFF

% ============================
\section{The X-ray light curve}
\label{sec:X-Ray-LC}
% ============================

Before we turn to analyse the column density evolution, we discuss the X-ray light curve shortly before, during, and right after the X-ray minimum. We emphasise three properties of the X-ray light curve, and explain them in the frame where the hard X-ray results from the post-shock secondary wind as it collides with the primary wind in the regions between the two stars (e.g., \citealt{PittardCorcoran2002, Akashietal2006}).
 
\textit{1. A deep X-ray minimum.} Figure \ref{fig:xray} shows that during the X-ray minimum the X-ray emission is very weak. Absorption alone cannot account for this. 
Hydrodynamic simulations by our group \citep{Akashietal2013, Kashi2017, Kashi2019} show that near periastron passage the secondary star accretes mass from the primary wind, a process that shuts down the secondary wind for several weeks.
For several weeks the flow of gas in the binary system is that of a secondary star accreting from the primary wind, rather than a flow of colliding winds.
These simulations reveal a complicated flow structure that is highly asymmetrical (apart from mirror-symmetry about the equatorial plane), and more complicated than the Bondi-Hoyle-Lyttleton (BHL; \citealt{HoyleLyttleton1939,BondiHoyle1944}) accretion picture. Nevertheless, we note the BHL accretion rate produces the accretion rate to within a factor of a few \citep{KashiSoker2009b}. The high $N_{\rm H}$ results in a lack of radiative cooling at the surface of the secondary star, further complicating the flow structure \citep{Akashietal2006}.
The \textit{NICER} light curve (Figure \ref{fig:xray}) suggests indeed that the secondary wind ceases to exist during the X-ray minimum. This seems to support the numerical results that the primary wind manages to reach the secondary star, and the secondary star accretes mass from the wind of the primary star rather than blowing its wind (e.g., \citealt{Soker2005, Kashi2019}; see section \ref{sec:intro}). According to this scenario, when the secondary star rebuilds its wind the X-ray emission resumes. In section \ref{sec:summary} we further discuss this point in relation to the orientation of the binary system.

During most of the X-ray minimum the X-ray emission is due mainly to post-shock secondary wind that was shocked weeks to months before periastron passage, and now resides at large distances of {$d_{x, {\rm min}} \ga 10 \AU$} from the primary star. 
By equation (\ref{eq:NHc}) or equation (\ref{eq:NHf}) this ensures a low column density, as we replace $r_{\rm{p}}$ with $d_{x, {\rm min}}$. 
Although the post-shock secondary wind suffers adiabatic cooling, its X-ray tail contributes to the $>5 \keV$ band. 
This explains the low values of $N_{\rm H}$ at the X-ray minimum.

\textit{2. An early exit from the X-ray minimum.} From Figure \ref{fig:xray_cycles} we learn that the last cycle had the earliest exit from the X-ray minimum. This continues the trend of the previous two cycles that had earlier exits than the first two cycles for which we have X-ray light curves \citep{Corcoran2005, Corcoranetal2017, Espinoza-Galeasetal2020}. As we mention in section \ref{sec:intro}, this might result from a weaker primary wind that allows the secondary wind to rebuild itself earlier. 
 
\textit{3. Strong pre-minimum fluctuations.} One feature common to all five light curves of the five cycles is that the light curve after exit from the minimum is relatively smooth. On the other hand, the light curves in the weeks to months before the minimum show large fluctuations, known as flares \citep{MoffatCorcoran2009}. This most likely is an outcome of large-amplitude instabilities in the process of the wind collisions, as numerical simulations show \citep{Akashietal2013, Kashi2017, Kashi2019}. This is important also for the minimum itself. The instabilities imply the presence of dense clumps in the primary wind. Such clumps are the first to reach the secondary star as the system approaches periastron, and they seem to weaken the secondary wind. This in turn allows more of the primary wind to reach the secondary star and completely or almost completely turn off the secondary wind for the duration of the minimum.

% ============================
\section{The column density to the hard X-ray source (5--10 \keV)}
\label{sec:NH}
% ============================

We perform our analysis under the common assumption that in $\eta$~Car most of the hard X-ray emission, $>5 \keV$, comes from the central binary \citep{Hamaguchietal2007}, namely from the wind collision region (specifically, the post-shock secondary wind). This region lies mainly between the two stars and is closer to the secondary star \citep[e.g.,][]{Ishibashietal1999, PittardCorcoran2002, Akashietal2006, Hamaguchietal2007}.
  
% ==============
\subsection{$N_{\rm H}$ close to apastron}
\label{sucsec:NHapastron}
% ==============

\cite{Hamaguchietal2007} deduced from their analysis that away from periastron (near apastron) the column density toward the hard X-ray emission component ($>5 \keV$) is $N_{\rm H,a} \simeq 17 \times 10^{22} \cm^{-2}$ (their figure 13). Subscripts `$a$' and `$p$' refer to values near apastron and near periastron, respectively. 

First we note that the X-ray light curve has large fluctuations (`flares' and `troughs'), as in previous cycles (e.g., \citealt{Corcoran2005}). The same goes for $N_{\rm H}$ toward the main hard X-ray source. From the left side of the upper panel of Figure \ref{fig:N_H} we find that away from periastron the values of the column density are in the range of $N_{\rm H,a} \simeq 20$--$60 \times 10^{22} \cm^{-2}$. 

\cite{KashiSoker2008a} calculated (their figure 6) the expected values of the absorbing column density for the two orientations close to apastron.
For $\omega \simeq 270^\circ$ (secondary closer to the observer at apastron) they obtained $N_{\rm H,a}{\rm (Cal270)} \simeq 4 \times 10^{22} \cm^{-2}$, while for $\omega \simeq 90^\circ$ (primary closer to the observer at apastron) they obtained $N_{\rm H,a}{\rm (Cal90)} \simeq 17 \times 10^{22} \cm^{-2}$.
It is clear that the $\omega \simeq 270^\circ$ orientation (secondary close to us near apastron) is not consistent with the high ($N_{\rm H,a} \simeq 20$--$60 \times 10^{22} \cm^{-2}$) column density derived in this work towards the wind collision region.
Therefore --- although there are large uncertainties and large variations from data point to data point --- overall, the new column density values we find from \textit{NICER} observations of the last cycle near apastron strengthen the claim of  \cite{KashiSoker2008a} for the $\omega \simeq 90^\circ$ orientation.   

% ==============
\subsection{$N_{\rm H}$ close to periastron}
\label{subsec:NHperiastron}
% ==============

To determine the column density toward the postshock secondary wind we consider only the hard X-rays, $>5 \keV$, that we attribute (see above) to the post-shock secondary wind, mainly close to the secondary star. 
During most of the $>5 \keV$ X-ray minimum, JD=2458889 (10-2-2020) to JD=2458912 (4-3-2020), the values of $N_{\rm H}$ are not much larger than the $N_{\rm H}$ values before and after the X-ray minimum, when the X-ray flux is much larger. Although the uncertainties in the values of $N_{\rm H}$ are on the order of the values themselves, the inferred change in $N_{\rm H}$ is sufficiently small to suggest that the minimum is \textit{not} due to absorption of the X-ray source. Hence, the X-ray source power must substantively diminish during the X-ray minimum.   

The general range of $N_{\rm H}$ values during the X-ray-minimum are in the range $28$--$72 \times 10^{22} \cm^{-2}$, with a median value of $\simeq 57 \times 10^{22} \cm^{-2}$.
Taking the weighted mean over this range we get $N_{\rm H,min}=(53 \pm 13) \times 10^{22} \cm^{-2}$.

We also note that when $\eta$ Car was observed during the the 2003.5 X-ray minimum both by \textit{XMM-Newton} and \textit{Chandra} (that has better spatial resolutions\footnote{\cite{Henleyetal2008} mention that the \textit{Chandra} observed spectral line profiles during the 2003.5 X-ray minimum can be fitted with synthetic profiles with a model of the emissivity along the colliding winds boundary.}), similar values of $N_{\rm H} \simeq 20$--$60 \times 10^{22} \cm^{-2}$ were derived from observations in both telescopes \citep{Hamaguchietal2007}.
This is also the same range of values we derive here during the 2020 X-ray minimum.

On JD=2458910 the binary system is about 20 days after periastron. At that time the binary orientation is at about $90^\circ$ degrees in the orbital motion with respect to periastron and the column density is about similar to the mean value during periastron.
There are few valid $N_{\rm H}$ observations at the exit from periastron so we will consider also values at later times, during the recovery (exit) from the minimum, JD=2458912 to JD=2458930. We note again that the exit from the minimum in the last cycle is the earliest among the five X-ray recorded cycles, and so the hard X-ray flux at exit is sufficiently strong to allow us determination of $N_{\rm H}$ at exit from minimum. During this exit period the values of the column density have a median of $\simeq 58 \times 10^{22} \cm^{-2}$ and the taking weighted mean over the period gives $N_{\rm H}{\rm (exit)}=(67\pm 11) \times 10^{22} \cm^{-2}$.
Thus we obtain
\begin{equation}
\frac{N_{\rm H,p}}{N_{+90^\circ}} \simeq 0.79 \pm 0.23,
\label{eq:ObservedRatio+}
\end{equation} 
where $N_{\rm H,p}$ and $N_{+90^\circ}$ are the means of the best-fit column densities during and after periastron passage.
  
Prior to the periastron passage the X-ray flux is larger, resulting in better determined values of $N_{\rm H}$.
About 20 days before the beginning of the X-ray minimum, JD $\simeq 2458870$, the binary system is about to enter the X-ray minimum. Taking the range JD=2458865 to JD=2458875 the column density has a median of $\simeq 25 \times 10^{22} \cm^{-2}$, and taking weighted mean over this range gives $N_{\rm H}{\rm (enter)} =(26 \pm 3) \times 10^{22} \cm^{-2}$. Adopting this value for the column density $90^\circ$ before periastron, i.e., $N_{\rm H}{\rm (enter)}=N_{-90}$, we find 
\begin{equation}
\frac{N_{\rm H,p}}{N_{-90}} \simeq 2.04 \pm 0.55,
\label{eq:ObservedRatio-}
\end{equation} 
where $N_{\rm H,p}$ and $N_{-90^\circ}$ are the means of the best-fit column densities during and before periastron passage.
 
Let us compare the observationally determined ratios in equations (\ref{eq:ObservedRatio+}) and (\ref{eq:ObservedRatio-}) with the theoretical expectation of equation (\ref{eq:ratios}). We first note that the simple theoretical setting we use in deriving equation (\ref{eq:ratios}) gives the same column density before and after the X-ray minimum, e.g., ${N_{-90}}={N_{+90}}$. Therefore, if we compare one value of the theoretical expectation with the observational findings it is the average of the values of equations (\ref{eq:ObservedRatio+}) and (\ref{eq:ObservedRatio-}), i.e., a ratio of $N_{\rm H,p}/N_{\rm H, 90} \simeq 1.4$ (recall that this is a theoretical ratio with about 1\% error).
For an assumed orientation in which the secondary star is closest to us at periastron ($\omega \simeq 90^\circ$) the theoretical ratio of the column density at periastron to that at $90^\circ$ orbit, which occurs about 20 days before and after periastron, is $N_{\rm H,c} /N_{\rm 90} \simeq 1.4$, while for the orientation where the secondary star is away from us at periastron ($\omega \simeq 270^\circ$) the expected value is $N_{\rm H,f} /N_{\rm 90} \simeq 3.3$ (equation \ref{eq:ratios}).
The observed ratios (equations (\ref{eq:ObservedRatio+}) and (\ref{eq:ObservedRatio-})) are hence more consistent with the $\omega=90^\circ$ orientation.

We note that it is possible that the $N_{\rm H}$ we deduce during the faintest portions of the \textit{NICER} X-ray minimum is not to the vicinity of the secondary star as we assume here, but rather to a different weak source near the center, what \cite{Hamaguchietal2007,Hamaguchietal2014a,Hamaguchietal2014b,Hamaguchietal2016} term the central constant emission (CCE) component.
We prefer the foregoing model --- wherein the intrinsic X-ray luminosity has declined (due to accretion onto the secondary) and $N_{\rm H}$ is indeed measured toward the central binary --- in part because of the smooth variation of $N_{\rm H}$ across the X-ray minimum. We would have expected that if the $N_{\rm H}$ during the X-ray minimum is measuring the absorbing column toward an X-ray source whose extent is much larger than the central binary, then $N_{\rm H}$ would show a steep drop as the flux drops when entering the X-ray minimum, and that these lower values are sustained until the flux recovers.
This signature is not apparent in these observations,
although we note that the value of $N_{\rm H}$ and its uncertainties are poorly constrained due to the low X-ray flux during minimum.
Similar smooth variation has been observed for the {1997--8} X-ray minimum by \textit{RXTE} \citep{Ishibashietal1999}.

% ==========================================================
\section{Summary and Discussion}
\label{sec:summary}
% ==========================================================

Beginning in July 2017, the \textit{NICER} X-ray telescope facility has regularly been observing $\eta$~Car, including daily coverage of the 2020.1 X-ray minimum that coincided with the so-called ``spectroscopic event'' of strong variability in visible and IR line emission which occurs around periastron passages (with last passage occurring in February 2020).
We processed and analyzed the \textit{NICER} X-ray observations (examples of which are shown in Figure \ref{fig:xrayspectra}), with our analysis focused on the hard (5--10 keV) X-ray light curve and the hydrogen column density to the source of the hard X-rays.
The light curve shows the expected X-ray minimum 5.54 years after the previous minimum in mid-2014. The \textit{NICER} data demonstrate that this most recent periastron passage exhibited the fastest recovery among the five observed X-ray minima (Fig \ref{fig:xray}).  

We interpret the fast recovery of this X-ray minimum in the frame of the accretion model of the spectroscopic event \citep{KashiSoker2008a}.
According to the accretion model, near periastron, the stellar wind collision region becomes very close to the secondary star; as a result, the secondary star accretes mass from the dense primary wind. This accretion process suppresses the secondary's wind.
We surmise that the primary wind was relatively weak during this most recent periastron passage, and this weak wind allowed the secondary wind to quickly revive itself, thereby terminating the X-ray minimum at the earliest recorded orbital phase after periastron passage.

By fitting the hard (5--10 keV) region of the X-ray spectra (see examples in Figure \ref{fig:xrayspectra510}), we determined the column density $N_{\rm H}(t)$ to the hard X-ray source as function of time (Figure \ref{fig:N_H}). The hard X-ray source originates close to the apex of the post-shock secondary wind, where the two winds collide directly, and is therefore located between the two stars, closer to the secondary than the primary star.
Although there are large uncertainties in the values of $N_{\rm H}(t)$ derived from the individual \textit{NICER} spectral fits, we found that we can use the time-averaged values of $N_{\rm H}$ during specific orbital phases (before, during, and just after the X-ray flux minimum) to discriminate between two alternative orientations proposed for the binary system (i.e., secondary in front of vs.\ behind primary during periastron passage). 
   
Specifically, in section \ref{sucsec:NHapastron}, we compared the values of $N_{\rm H}$ that we derived from observations away from periastron, i.e., near apastron (left region of upper panel of Figure \ref{fig:N_H}) with the theoretically expected values that we derived in \cite{KashiSoker2008a}. We found that the column densities away from periastron are too large to be consistent with the binary orientation wherein the secondary star is closer to us at apastron, i.e., $\omega \simeq 270^\circ$, since we require the primary dense wind to supply the column density. These results thereby support the earlier conclusion of \cite{KashiSoker2008a} \citep[which were based on $N_{\rm H}$ values derived from \textit{XMM-Newton} X-ray observations;][]{Hamaguchietal2007} that the secondary star of $\eta$~Car is away from us near apastron, i.e., $\omega \simeq 90^\circ$.  

In section \ref{subsec:NHperiastron} we compared the values of $N_{\rm H}$ that we calculated for the two alternative opposite binary orientations (section \ref{sec:theory_NH}) to those that we derived from observations (lower panel of Figure \ref{fig:N_H}). 
We took the periastron passage to have occurred just after the beginning of the X-ray minimum. We determined the approximate ratio of the column density near periastron to that about 20 days later, when the primary-secondary position had changed by $90^\circ$ (equation \ref{eq:ObservedRatio+}), and the ratio of periastron column density to that 20 days before periastron (equation \ref{eq:ObservedRatio-}). These ratios of $N_{\rm H}$ variation are $\approx 0.8$ and $\approx 2.0$, respectively. 
In the simple theoretical model we have developed (Figure \ref{fig:Schematic}), these two ratios are aught to be small if $\omega \simeq 90^\circ$, since during this time the line of sight to the hard X-ray source should not reach very close to the primary where the densities are very large.
This range of column density ratio variation, 0.8--2.0, can then be compared with the theoretical predictions for the variations that should result from the two opposing assuming binary orientations. For the $\omega = 90^\circ$ orientation (upper part of Figure \ref{fig:Schematic}) the predicted ratio of variation is 1.4 --- precisely in the middle of the observed range --- while for the $\omega = 270^\circ$ orientation (lower part of Figure \ref{fig:Schematic}) the predicted ratio of variation is 3.3 (equation \ref{eq:ratios}), i.e., outside the range of observed variation.
Therefore, this comparison of the theoretically expected variation of $N_{\rm H}$ around periastron (that we express as ratios with the value at the X-ray minimum) to those that we derived from observations, also supports the $\omega=90^\circ$ orientation (upper part of Figure \ref{fig:Schematic}).

Thus, the main conclusion we draw from our analysis of the \textit{NICER} X-ray observations of the most recent periastron passage of $\eta$~Car is that the binary orbital orientation is $\omega \approx 90^\circ$, i.e., the secondary star is closer to us than the primary star at periastron.
Our secondary conclusion is that the weakening of the primary stellar wind over the last several cycles allowed the earlier revival of the secondary wind and, resulting in the fastest recovery from X-ray minimum yet observed for $\eta$~Car.

The results described in this paper serve to illustrate how the variation of column density toward the hard X-ray source during the orbital motion of the massive binary system $\eta$~Car, $N_{\rm H}(t)$, is potentially a more sensitive diagnostic of the configuration of the binary and the orientation of the binary orbit than is the variation of the X-ray flux $F_{\rm X}(t)$ \citep{KashiSoker2008a}. The drawback of this application of column densities derived from spectral fitting is that --- especially in the (hard X-ray) energy range of interest here ($\sim$5--10 keV) --- $N_{\rm H}(t)$ is subject to larger uncertainties than $F_{\rm X}(t)$. Thus, in the coming decades, higher quality X-ray observations of $\eta$~Car around periastron utilizing planned X-ray missions such as Athena+ \citep{Nandraetal2013} are essential if we are to improve our understanding of this astrophysically important massive binary system.

\vspace{0.5cm}
We thank an anonymous referee for helpful comments. This paper used data from the Neutron star Interior Composition Explorer (\textit{NICER}), obtained from the High Energy Astrophysics Archive Research Center (HEASARC).
We thank R. Remillard for his helpful discussion regarding use of the \textit{NICER} \texttt{nibackgen3c50} tool.
AK acknowledges support from the R\&D Authority, and the chairman of the Department of Physics in Ariel University.
NS was supported by a grant from the Israel Science Foundation (769/20).
\software{Sherpa, (v4.12; \citealt{Freemanetal2001,Doeetal2007,Burkeetal2020}, HEAsoft (v6.27.2; HEASARC 2014))}
\software{nicerl2, HEAsoft (v6.27.2; HEASARC 2014), \url{https://heasarc.gsfc.nasa.gov/lheasoft/ftools/headas/nicerl2.html}}
\software{nibackgen3c50, HEAsoft (v6.27.2; HEASARC 2014), \url{https://heasarc.gsfc.nasa.gov/docs/nicer/tools/nicer_bkg_est_tools.html}}
\software{VAPEC (\citealt{MorrisonMcCammon1983}; HEAsoft (v6.27.2; HEASARC 2014)),
\url{https://heasarc.gsfc.nasa.gov/xanadu/xspec/manual/node135.html\#vapec}}

\end{document}